\documentclass[onecollarge]{svjour2}       
\smartqed  
\usepackage{graphicx}
\usepackage{dcolumn}
\usepackage{bm}
\usepackage{amsfonts}
\usepackage{amsmath}
\usepackage{amssymb}
\usepackage{color}
\usepackage{float}

\usepackage{changes}
\definechangesauthor[name={E}, color=orange]{E}

\journalname{J. Stat. Phys.}

\begin{document}

\newcommand{\be}{\begin{equation}}
\newcommand{\ee}{\end{equation}}
\newcommand{\bea}{\begin{eqnarray}}
\newcommand{\eea}{\end{eqnarray}}
\newcommand{\bse}{\begin{subequations}}
\newcommand{\ese}{\end{subequations}}
\newcommand{\comment}[1]{}
\newcommand{\eps}{\varepsilon}

\newcommand{\Dt}{{\Delta t}}
\newcommand{\Dx}{{\Delta x}}
\newcommand{\Dv}{{\Delta v}}
\newcommand{\DW}{{\Delta W}}
\newcommand{\nl}{\newline}
\newcommand{\mean}[1]{{\left \langle {#1} \right \rangle}}
\newcommand{\green}[1]{\textcolor{green}{#1}}
\newcommand{\red}[1]{\textcolor{red}{#1}}
\newcommand{\nota}[1]{\textcolor{red}{\it (#1)}}

\renewcommand{\ss}[1]{_{\hbox{\tiny #1}}}
\newcommand{\us}[1]{^{\hbox{\tiny #1}}}

\def\bfi{\begin{figure}[H]}
\def\efi{\end{figure}}

\title{One dimensional phase-ordering in the Ising model with space decaying
interactions}


\author{Federico Corberi \and Eugenio Lippiello \and Paolo~Politi}

\institute{ Federico Corberi
           \at  Dipartimento di Fisica ``E.~R. Caianiello'', and INFN, Gruppo Collegato di Salerno, and CNISM, Unit\`a di Salerno,Universit\`a  di Salerno,
via Giovanni Paolo II 132, 84084 Fisciano (SA), Italy\\
	   \email{corberi@sa.infn.it}
	   \and
	   Eugenio Lippiello
           \at Dipartimento di Matematica e Fisica, Universit\`a della Campania,
Viale Lioncoln 5, 81100, Caserta, Italy\\
	   \email{eugenio.lippiello@unicampania.it}
	   \and
           Paolo Politi \at
           Istituto dei Sistemi Complessi, Consiglio Nazionale
           delle Ricerche, via Madonna del Piano 10, I-50019 Sesto Fiorentino, Italy\\
           INFN Sezione di Firenze, 
	   via G. Sansone 1 I-50019, Sesto Fiorentino, Italy\\
           \email{Paolo.Politi@isc.cnr.it}
}

\date{Received: date / Accepted: date}

\maketitle

\begin{abstract}
The study of the phase ordering kinetics of the ferromagnetic one-dimensional Ising model dates back to 
1963 (R.J. Glauber, J. Math. Phys. 4, 294) for non conserved order parameter (NCOP)
and to 1991 (S.J. Cornell, K. Kaski and R.B. Stinchcombe, Phys. Rev. B 44, 12263) for conserved order parameter (COP).
The case of long range interactions $J(r)$ has been widely studied at equilibrium but their effect on relaxation
is a much less investigated field. 
Here we make a detailed numerical and analytical study of both cases, NCOP and COP.
Many results are valid for any positive, decreasing coupling $J(r)$, but 
we focus specifically on the exponential case,
$J\ss{exp}(r)=e^{-r/R}$ with varying $R>0$, and on the integrable power law case, $J\ss{pow}(r) =1/r^{1+\sigma}$ with $\sigma > 0$.
We find that the {\it asymptotic} growth law $L(t)$ is the usual algebraic one, $L(t)\sim t^{1/z}$, 
of the corresponding model with nearest neighborg interaction ($z\ss{NCOP}=2$ and $z\ss{COP}=3$)
for all models except $J\ss{pow}$ for small $\sigma$: in the non conserved case when $\sigma\le 1$ ($z\ss{NCOP}=\sigma+1$)
and in the conserved case when $\sigma\to 0^+$ ($z\ss{COP}=4\beta+3$, where $\beta=1/T$ is the inverse of the absolute temperature).
The models with space decaying interactions also differ markedly
from the ones with nearest neighbors due to the presence of many long-lasting
preasymptotic regimes, such as an exponential mean-field behavior with $L(t)\sim e^{t}$,
a ballistic one with $L(t)\sim t$, a slow (logarithmic) behavior
$L(t)\sim \ln t$ and one with $L(t)\sim t^{1/\sigma+1}$.
All these regimes and their validity ranges have been found analytically and verified in numerical simulations. 
Our results show that the
main effect of the conservation law is a strong slowdown of COP dynamics
if interactions have an extended range.
Finally, by comparing
the Ising model at hand with continuum approaches based
on a Ginzburg-Landau free energy, we discuss when and to which extent
the latter represent a faithful description of the former.

\end{abstract}


\section{Introduction}

Phase ordering~\cite{Bray94} is the dynamical process of growth of order when a system is quenched from a high temperature homogeneous phase
to a low temperature broken-symmetry phase. It occurs through domain coarsening, with the average size of domains of
different phases, $L(t)$, which increases in time.

Phase ordering is an old research topic and its most unified picture is based on a continuum approach
whose starting point is a Landau-Ginzburg free energy.
In an impressive series of papers dating back around twenty five years, Alan Bray and collaborators 
have constructed a theory which covers nonconserved and conserved models,
scalar and vector fields, short and long range interactions~\cite{Bray94,BrayRut94,RutBray94}.
Their results for the growth law $L(t)$ do not depend on the spatial dimension $d$ of the physical system, which only
appears when defining the limits of applicability of the theory.

It is clear that one dimension, the case we focus on in this paper, plays a special role for scalar systems because the Curie temperature vanishes 
for short-range interactions, $T_c=0$. This means, first of all, that it is not possible to quench the temperature
from $T_i > T_c$ to $T_f\equiv T < T_c$. However, we can consider quenches to a
vanishing $T=0$, or to a very small finite temperature
$T \ll 1$ (the energy scale of coupling is of order one and here and in the following we set the Boltzmann constant to unity).
In the former case ($T=0$) the equilibrium state is 
fully ordered and the dynamics can never increase the system energy.
This means that zero temperature dynamics can be blocked, which is actually what happens if the order parameter is
conserved.
In the latter case ($T>0$) the final equilibrium state
is made of ordered regions of average size equal to the equilibrium correlation length $\xi$ 
and, hence, coarsening stops when $L(t) \simeq \xi(T)$. 
Since $\xi$ is a very fast increasing function of $(1/T)$ (for nearest neighbor (nn) interactions, $\xi(T)\simeq e^{2/T}$), 
low$-T$ coarsening dynamics lasts for a long time.

The nn Ising model has been studied decades ago. Nonconserved
dynamics proceeds via spin-flips and it is possible to attain
the ground state through processes which lower or keep constant the energy. We will see that dynamics can be easily
described in terms of random walks performed by domain walls, so it is not surprising that the average size of domains grows
according to the law~\cite{Glauber1963} $L(t) \simeq t^{1/2}$.
Conserved dynamics proceeds differently because single spin-flips, which would change
the order parameter (the magnetization), are not allowed.
In this case we rather have spin-exchange processes with spins
that can {\it evaporate} from a domain wall and condensate to another droplet
after a diffusion process. It is known that such evaporation-condensation mechanism slows down 
the dynamics~\cite{Cornell1991} with respect to spin-flip and results in the growth law
$L(t)\simeq t^{1/3}$. 

In a recent publication~\cite{EPL} we have studied the effects of
a coupling constant $J(r)=e^{-r/R}$, decreasing exponentially with the distance
$r$ between two spins, on the one-dimensional coarsening dynamics.
Since this interaction introduces the new length scale $R$, it is reasonable to expect
the regime $L(t) < R$ to be physically different from the one with $L(t) > R$. 
We actually found more than that, because we identified different dynamical regimes for large $L(t)$.
In this paper we go beyond the exponential coupling, finding a series 
of results which are independent of the explicit
form of the coupling, provided that $J(r)$ is a positive, decreasing function of $r$.
For definiteness, detailed simulations and specific calculations have been done for
exponential, $J\ss{exp}=e^{-r/R}$, and power-law, $J\ss{pow}(r)=1/r^{1+\sigma}$, couplings. 
We focus on the growth law $L(t)$ of the domains' size, which we compute
by means of different analytical approaches along the whole time history,
from the instant of the quench up to the asymptotic stages. Our results are
successfully compared to the outcome of numerical simulations.

Besides addressing the modifications of the kinetics due to a space decaying interaction,
in this paper we also discuss an interesting question which has not been considered previously, namely the comparison between
the coarsening dynamics of the Ising model and the one emerging from a
deterministic continuum description of the same system.
This analysis allows us to provide a physical interpretation
to the various dynamical regimes and to show that some of them,
although occurring with the same coarsening law in the discrete
and in the continuum models, are associated to different physical mechanisms.
We will also comment on a recent preprint~\cite{1901.01756} studying phase ordering for
a strictly related Non Conserved Order Parameter (NCOP) long-range discrete model.

This paper is organized as follows: in Sec.~\ref{models}
we introduce the models, we define the form of the interactions and discuss the 
equilibrium structure. We also specify the kinetic rules and
discuss the elementary processes driving the evolution and the methods that can be used to
perform numerical simulations. 
Section \ref{simplified} is devoted to the definition of the simplified models
with only few domains from which most of our analytical results can be
deduced. In Sec.~\ref{sec.NCOP} we focus on the NCOP case, by deriving our
analytical predictions and comparing them with the outcome of numerical
simulations. We do the same in Sec.~\ref{sec.COP} for the Conserved Order Parameter (COP) case.
In Sec.~\ref{sec_continuum} we compare our results for the discrete Ising
model with the behavior of the continuum model based on a Ginzburg-Landau
free energy, discussing to which extent the latter can grasp the physics
observed in the former. In Sec.~\ref{concl} we discuss the results
of this paper on general grounds, and suggest some possible future research lines, while
in the Appendix we give some details of numerical simulations.
The captions of most figures showing our results are preceded by an abbreviation
to immediately identify the model in question. For example,
{\bf (NCOP exp)} means we are considering the exponential model of the nonconserved class.

\section{The models} 
\label{models}

We will consider a general one-dimensional Ising model described by the Hamiltonian 
\be
{\cal H} = -\sum_{i=-\infty}^{+\infty} \sum_{r>0} J(r) s_i s_{i+r} ,
\label{HIsing}
\ee 
where $s_i=\pm 1$ are binary variables and
we only assume that $J(r)$ is a positive, decreasing function of the distance 
$r$.

Main formulas will be written for general $J(r)$, but specific calculations and simulations will
be done for an exponentially decreasing coupling,
\be
J\ss{exp}(r) = e^{-r/R}
\label{J_exp}
\ee
and for a power-law one,
\be
J\ss{pow}(r) = \frac{1}{r^{1+\sigma}} ,
\label{J_power}
\ee
with $\sigma>0$.

The usual Ising Hamiltonian couples only nearest neighbour spins,
\be
J\ss{nn}(r)=\delta_{r,1}.
\label{J_nn}
\ee

In all cases, Eqs.~(\ref{J_exp}-\ref{J_nn}), $J(r)$ should be proportional to some energy scale $J_0$, which will be assumed
to be equal to one throughout all the paper. 

Since we will use both a spin and a lattice gas language, it is useful to rephrase the Ising
Hamiltonian (\ref{HIsing}) using the variable $n_i = (1+s_i)/2$, which takes the values $n_i=1$ (for $s_i=+1$) 
and $n_i=0$ (for $s_i=-1$):
\bea
{\cal H} &=& -\sum_{i=-\infty}^{+\infty} \sum_{r>0} J(r) (2n_i -1)(2n_{i+r} -1) \\
&=& -4\sum_{i=-\infty}^{+\infty} \sum_{r>0} J(r) n_i n_{i+r} 
+ 4 \sum_{i=-\infty}^{+\infty} n_i \sum_{r>0} J(r)
-  \sum_{i=-\infty}^{+\infty} \sum_{r>0} J(r) \; .
\label{energygaslat}
\eea
The third term on the right-hand side is an irrelevant (extensive) constant.
The second term on the right-hand side 
is constant as well if the dynamics preserves the order parameter, i.e. if $\sum_i s_i =\sum_i (2n_i - 1)$
is a constant of motion. The spin language is appropriate for NCOP systems,
like magnetic systems; the lattice gas language is appropriate instead for COP
systems, like alloys or fluids. Within this language, the coupling energy between 
two particles at distance $r$, see Eq.~(\ref{energygaslat}), is $-4J(r)$.

As for the equilibrium properties, it is straightforward that there are no qualitative
differences between $J\ss{exp}(r)$ and $J\ss{nn}(r)$. 
The power-law case, instead, requires a few more words. If GS is one of the two ground states and we flip one single spin obtaining
the (one flip) state 1F, we have the energy difference
\be
E\ss{1F} - E\ss{GS} = 4 \sum_{r=1}^{\infty} \frac{1}{r^{1+\sigma}} ,
\ee
which diverges for $\sigma \le 0$. For negative $\sigma$ we are in the so called {\it strong} long-range regime,
where extensivity and (above all) additivity do not hold. We will not consider this case,
which is reviewed in Ref.~\cite{review_long_range}. 

If we flip all spins $i>0$, therefore obtaining the microscopic state 1DW characterized by a single domain wall,
we have the energy difference
\be
E\ss{1DW} - E\ss{GS} = 2  \sum_{i=1}^{\infty} \sum_{j=-\infty}^0 \frac{1}{(i-j)^{1+\sigma}} ,
\ee
which is finite for $\sigma > 1$. Therefore 
the standard arguments to explain the absence of long-range~\cite{Peierls1934,Dyson1969}
order at finite temperature still applies. For $0 < \sigma \le 1$ ({\it weak}
long-range regime) there is an ordered phase at finite temperature:
for $0< \sigma < 1$ there is a second-order phase transition~\cite{Dyson1969} while
for $\sigma=1$ there is a Kosterlitz-Thouless phase transition with a jump of the 
magnetization ~\cite{Frohlich1982,Imbrie1988,Luijten2001}.
It is also possible to argue that mean-field critical exponents are expected for 
$0<\sigma <\frac{1}{2}$~\cite{Mukamel2009}.

\subsection{Dynamical evolution}

All models will be studied for both nonconserved and conserved order parameter.
The simplest NCOP dynamics is based on single spin-flip processes, also called Glauber dynamics: $s_i \to -s_i$.
Instead, the simplest COP dynamics is implemented by a spin-exchange processes between two opposite spins, also called
Kawasaki dynamics: $s_i \leftrightarrow s_j$. In the lattice gas language, if (say)
$s_i=1$, this move is simply the hopping
of a particle on site $i$ towards an empty site $j$: $(n_i,n_j) =(1,0) \to (0,1)$. In most cases and in this
manuscript as well, sites $(i,j)$ must be nn: $j=i\pm 1$. 

The transition rates $W\ss{IF}$ between an initial state $I$ and a final state $F$ must
satisfy detailed balance in order to ensure relaxation to equilibrium, $W\ss{IF} / W\ss{FI} = e^{-\beta (E_{F} -E_{I})}$ in a canonical ensemble. 
If we impose the additional constraint $W\ss{IF} + W\ss{FI} = 1$ we obtain the Glauber transition rates,

\be
W\ss{IF} = \frac{1}{1+e^{\beta(E_{F} -E_{I})}} .
\label{glauber}
\ee
These transition rates will be used in the following for the elementary moves, namely
spin flips for NCOP or spin exchanges for COP.

Starting with the NCOP case, in Fig.~\ref{fig.moves}(a) we plot the three possible different single spin-flip processes.
We use here also a language which associates a domain wall (DW) to a pair of nn antiparallel spins.
Within such language the three processes correspond to diffusion, annihilation and creation of DWs.

It is
important to evaluate for each process the energy difference $\Delta E = E_{F} - E_{I}$ between the final and the initial state. 
For a nn Ising model, it is straightforward to conclude that $(\Delta E)\ss{nn} =0$ for DW diffusion, $(\Delta E)\ss{nn}<0$ for annihilation
of two DWs, and $(\Delta E)\ss{nn}>0$ for the creation of two DWs.
If $J(r) \ne 0$ for $r>1$ it is not possible to make general statements
about the sign of $\Delta E$ for any pair of initial and final configurations differing for a single spin flip, since this task involves the knowledge of the whole system configuration.
However, as argued in Sec.~\ref{sec.NCOP}, most of NCOP dynamics can be understood making reference to 
a simple configuration, see Fig.~\ref{models}(a). In this case we still have $\Delta E<0$ for DW annihilation
and $\Delta E>0$ for DW creation, but DW hopping is no longer free diffusion, because $\Delta E \ne 0$ and its
sign is such that $W\ss{IF}$ favors the closing of the smaller domain.
Dynamics at $T=0$ is therefore trivial:
DWs cannot be created and each DW drifts to the closest DW until annihilation occurs. When temperature is switched on, two
things happen: DW drift becomes an asymmetric DW diffusion and DW creation is permitted.
Asymmetric diffusion makes dynamics much more complicated because the symmetric and the antisymmetric parts of DW hopping compete
and their balance depends on the temperature, on how $J(r)$ decreases with $r$, and on the typical distance $L(t)$ 
between neighbouring DWs.
DW creation at finite $T$ makes possible to attain thermal equilibrium, with a distance $L(t)$ between DWs which
equals the equilibrium correlation length, $\xi(T)$, similarly to what happens in the nn case.

Let us now move to the COP class, whose
elementary processes at the basis of the dynamics are shown in Fig.~\ref{fig.moves}(b).
We use again a particle language where now a particle represents $n_i=1$ while an empty sites (or a hole)
corresponds to $n_i=0$. The four elementary processes are particle and hole diffusion, particle attachment and
particle detachment. If we limit to nn interaction it is obvious that diffusion keeps the energy constant, 
attachment decreases the energy and detachment increases the energy.
Going beyond nn interaction makes impossible general statements about the sign of $\Delta E = E_{F} - E_{I}$,
as in the NCOP case, but, again, if we focus on the simple configuration
depicted in Fig.~\ref{models}(b) we can make some statements: detachment costs energy, attachment gains energy
and diffusion is asymmetric with a drift towards the closest domain.~\footnote{More precisely, the drift points
to the closest domain if the two domains are of equal length. This means that the direction
of the drift changes when the diffusing particle reaches the middlepoint between the
two domains. In the general case of unequal domains,
the drift changes direction when the diffusing particle moves from one domain to the other, but not in the middle point.}

It is clear that for COP, temperature has a more direct role in the dynamics, which can be easily highlighted in the nn model.
At $T=0$ the allowed processes are particle and hole diffusion, and particle attachment.
This means that the dynamics stops as soon as a particle is attached to another particle and a hole is attached to
another hole. Starting from a fully disordered configuration, $T_i=\infty$, we obtain an average domain size 
$L\ss{nn}(T=0)\simeq 4.135$~\cite{L_asint}. 
If $J(r)$ extends beyond nearest neighbours, being attached to another particle/hole is not enought to avoid diffusion
but for not so small domains it is surely true that detachment costs energy and it is forbidden at $T=0$.
We can therefore affirm that zero-temperature dynamics stops almost immediately, with a ``small" average size of
domains.
When we switch on temperature, particle detachment is allowed.
Monomers perform an asymmetric diffusion and they can travel the whole empty space between
two clusters and reach the other domain. 
The probability that this occurs will be evaluated in Sec.~\ref{sec.COP}.
This is the mechanism whereby neighbouring domains exchange monomers and represents
the basic process leading to coarsening since, due to the loss or gain of monomers, the length of each
domain performs a sort of random walk. 

If a second particle is detached in the empty domain between two clusters while the first one
is still diffusing (an extremely rare event at low $T$)
they can stick together forming a new cluster. This is the mechanism arresting coarsening and producing equilibration 
when $L(t) \simeq \xi(T)$.

\bfi
\begin{center}
\includegraphics*[width=0.95\columnwidth,clip=true]{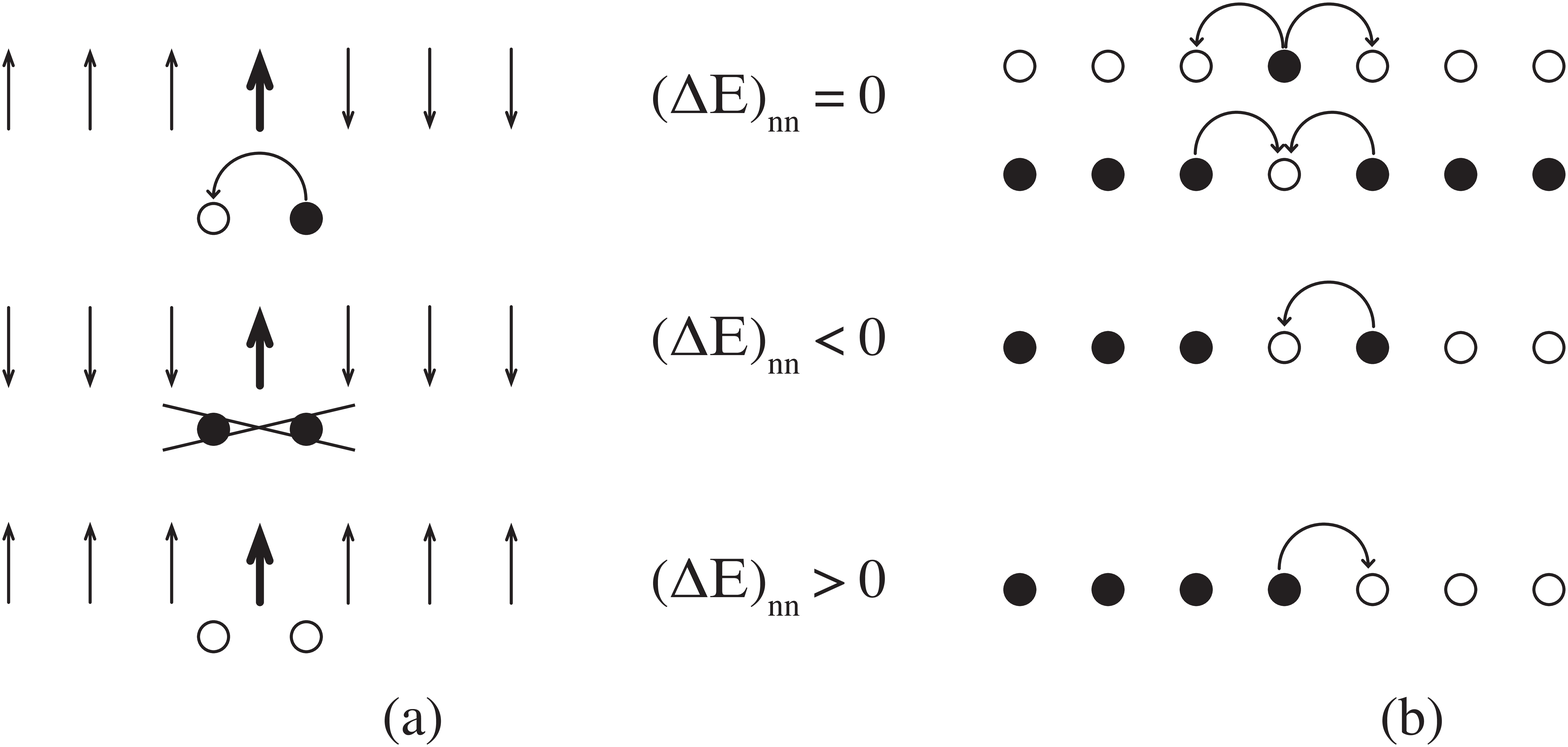}
\end{center}
\caption{(a) NCOP (Glauber) dynamics, using spin language and domain wall (DW) language
  (a DW is represented by a circle). The reversed spin is the thick spin
and its flipping implies the hopping of a DW (top), the annihilation of two DWs (center), or the creation of two DWs (bottom).
DWs are solid and open circles, before and after the flipping, respectively.
(b) COP (Kawasaki) dynamics, using the lattice gas language. Full circles are particles, empty circles are holes.
Arrows indicate possible moves of a particle.
NCOP and COP microscopic processes are classified according to the variation of energy for the nn Ising model,
$(\Delta E)\ss{nn} =E_F - E_I$, between the final and the initial state.}
\label{fig.moves}
\efi

\subsection{Numerical simulations} \label{simuldes}

We consider  a chain of $N$ spins with periodic boundary conditions and implement standard Monte Carlo dynamics with
Glauber transition rates, see Eq.~(\ref{glauber}). Since the interaction extends to all the spins,
for each spin flip trial, the evaluation of the energy cost $\Delta E = E_{F} - E_{I}$ involves the sum over all spins in the system with a computation time $N \times N$. This makes  simulations much slower than in the nn case and
prevents one to obtain completely satisfactory
results in some cases. An efficient algorithm has been developed in 
Ref.~\cite{met_num_LR} which has shown that, in the case of periodic boundary conditions,
it is possible to obtain an efficient diagonalization of the coupling matrix $J(r)$ via FFT. By means of this method, one obtains 
an exact algorithm that scales with $N \times \ln N$. In our study we consider two  different approximated 
simulation schemes which scale with $N$ and with the number of defects $n$, respectively. The two schemes are
briefly introduced below  and a more detailed discussion on simulations techniques is contained in Appendix \ref{Appendix}.

\begin{itemize}

\item[S1] {\it Simulations with truncated $J(r)$.}

  In this case we consider the Hamiltonian (\ref{HIsing}) in terms of the spin variables and assume, for  exponential couplings $J\ss{exp}(r)$, that  $J(r)\equiv 0$ when $r>{\cal M}R$,
  where ${\cal M}$ is a sufficiently large number. We have checked that with
  ${\cal M}$ of order $10^2$ the results obtained with truncation are
  indistinguishable from exact simulations. With such large
  values of ${\cal M}$, therefore, this kind of simulations is basically
  exact.
\item[S2] {\it Simulations with a reduced number of interacting kinks.}

This method implements the DW description illustrated 
in Fig.(\ref{fig.moves}) where a spin flip or a pair of spin flips is mapped in different moves of a {\it particle}. 
The key observation is that, indicating with $n$ the number of DWs in the system at a given time, the energy 
difference $\Delta E$ after the flip of the spin in the $i$-th site can be always written in the form (see Appendix \ref{Appendix})
\be
\Delta E= (\Delta E)_{nn} + \sum_{\substack{j=-n/2 \\x_j \neq i}}^{n/2} (-1)^j Q\left(\vert x_j -i\right\vert) , 
\label{deltaEii}
\ee
where $x_k$ is the position of the $k$-th DW and DWs are sorted according to their distance from the site $i$.   
In the above equation $(\Delta E)_{nn}$ is the quantity specified in Fig.~\ref{fig.moves}, namely 
the nn contribution to the energy change for the different moves.  $Q(r)$ is a decreasing function of $r$, 
proportional to $J(r)$ or to its integral for COP and NCOP dynamics, respectively. 
The numerical implementation of this simulation method is more complex but the evaluation of $\Delta E$ 
involves the sum of $n$ terms, while method S1 involves the sum of ${\cal M}R$ terms, therefore making S2 
particularly advantageous at large times, when $n = N/L(t) \ll {\cal M} R$. 
On the other hand, at short times $n$ is comparable to the total number of spins in the system and S2 becomes less efficient. 

Within the exact S2 framework 
one can introduce an approximated simulation method corresponding to consider an interaction extending only to a finite number of 
kinks $n_K$, by means of the substitution $n\to n_K$ in Eq.~(\ref{deltaEii}), where $n_K$ is a parameter
to be optimized. 
Clearly, the smaller is $n_K$ the worst, but the faster, the approximation is.
This approximation is expected to provide exact results for sufficiently large times 
when $Q(\vert x_j-i\vert)$ can be neglected for $j>n_K$.
In the following we will refer to S2 assuming a suitable $n_K$. 

\end{itemize}

In the case of COP dynamics we always use the simulation method S2 
whereas for NCOP we use S1 for simulations at short times and S2 for longer simulations. 
In addition, for NCOP, to speed further up the computations, 
we have considered simulations where activated processes (spin-flips in the bulk) 
are forbidden, preventing the equilibration of the system. This approximation becomes exact in the limit of 
very small temperature and at large $t$, when the average distance between domains is sufficiently large 
so that the sum in Eq.(\ref{deltaEii}) becomes much smaller than $2 J(1)$.  
For COP, instead, we adopt a rejection free algorithm 
of the type described in~\cite{Bortz75}
where activated moves (such as monomers evaporations), which in the low $T$
limit are very unlikely and delay the dynamics, are always accepted and time is 
increased according to the likeliness of the accepted event (see Appendix \ref{Appendix}). 
This technique does not introduce any error.

In our simulations we have consider a chain of $N=10^4-10^8$ spins, depending on the various parameters.
These numbers are sufficiently large to avoid
finite-size effects in the range of time considered.
For each choice of the parameters, we take an average over $10^2-10^3$ realizations
of the initial conditions and of the thermal history.
Additional details of the used algorithms can be found in Appendix~\ref{Appendix}. 

We finally stress that for the models which are magnetized at finite temperature
(power models with $\sigma < 1$) we always consider a quench to the ordered phase.
In particular, for $\sigma=0.5$ we take $T\le 1$ while $T_c > 4$~\cite{Tc_sigma}.

\section{Simplified models with few domains} \label{simplified}

After a quench from the fully disordered phase ($T_i=\infty$) to zero or low temperature
the system relaxes to equilibrium, which is characterized by a local order
on a length scale equal to the correlation length $\xi(T)$. 
Upon increasing the interaction range, the correlation length increases as well
and for the class of power law models we even have long range order at finite $T$, if
$\sigma \le 1$. In any case, even if $T\ss{c}=0$ we can consider low enough $T$ such that $\xi(T)$
is arbitrarily large.

\bfi
\begin{center}
\includegraphics*[width=0.95\columnwidth,clip=true]{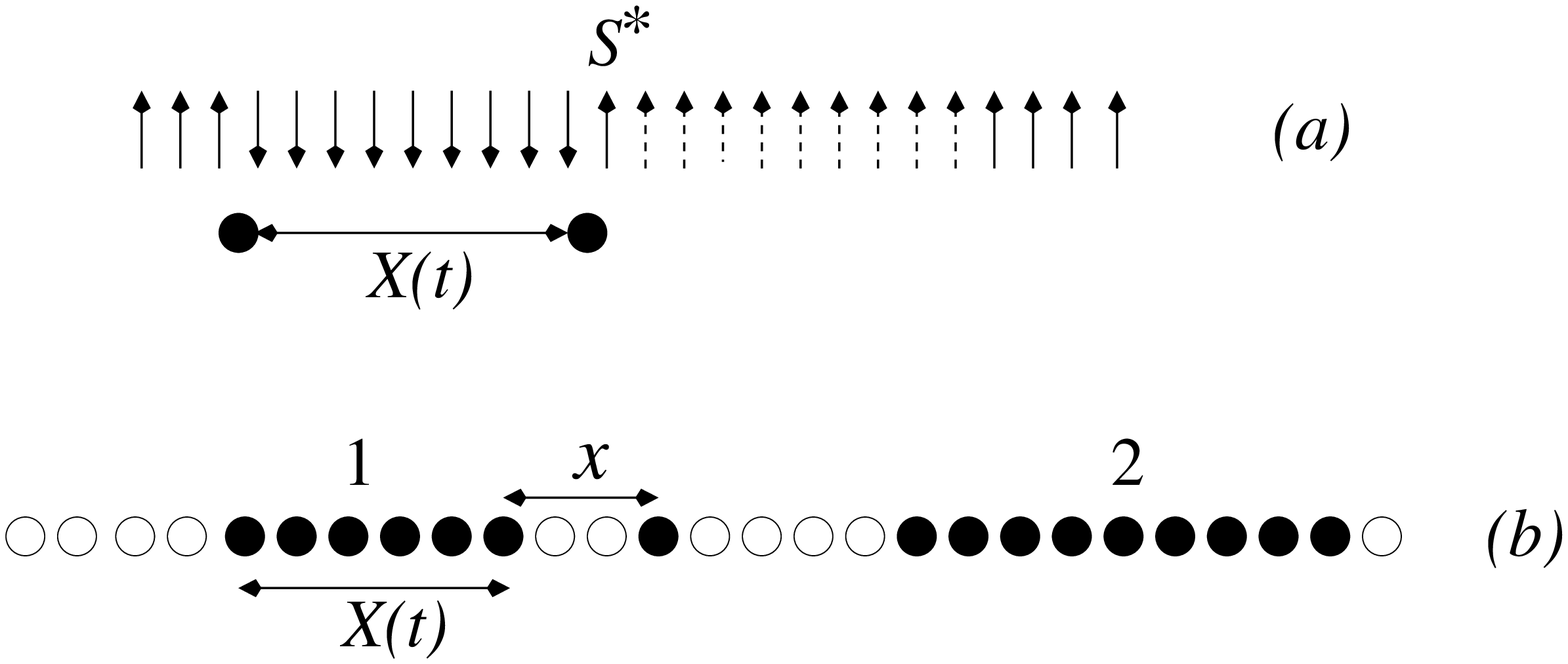}
\end{center}
\caption{
Simple one dimensional configurations, with periodic boundary conditions.
(a) A single domain of down spins and length $X(t)$ (with $X(0)=L$) and a neighbouring domain 
of initial length $L_2$ (which scales with $L$). 
The interaction of spin $S^*$ with down spins is compensated by its
interaction with dashed up spins.
(b) Two clusters of particles, labelled as ``1" and ``2", whose sizes change in time because they exchange monomers.
The distance between the monomer and cluster 1 is $x$.
}
\label{fig.models}
\efi

Relaxation to equilibrium occurs through a coarsening kinetics where domains disappear leading to 
the increase over time of their average size, $L(t)$.
Figure~\ref{fig.models} represents the simplest configurations leading to the disappearance of a single domain, for NCOP 
(a) and COP (b).
The analysis of the dynamics of such configurations not only explains the physics of the process, 
but it also allows to derive most of the coarsening laws.
In fact, the evolution is a self-similar phenomenon characterized by a single length scale, $L(t)$.
According to the scaling hypothesis, this property means, e.g., that the correlation function
$C(r,t) = \langle s_i(t) s_{i+r}(t)\rangle$
is actually a function of a single variable, $C(r,t) = f(r/L(t))$.
Invoking scaling, we argue that the functional dependence $L(t)$ can be found 
by determining the typical temporal scale $t$ necessary to close a domain of initial size $L$
and inverting the resulting function $t(L)$. 
Let us now see how this program can be implemented.

Starting from the NCOP case, see Fig.~\ref{fig.models}(a), we consider a domain of the negative phase
and initial size $X(0)=L$ and a neighbouring domain of the positive phase
and initial size $L_2\ge L$, which scales with $L$ (periodic boundary conditions apply). 
Each DW performs an anisotropic random walk with a drift which favors the closing of the smallest domain,
whose closure time is $t(L)$.
In the next Section we will evaluate the drift in the configuration $L_2=\infty$ and the closure time
in the configuration $L_2=L$. This choice is due to the fact that the drift vanishes for $L_2=L$ 
and the closure time may diverge for $L_2=\infty$.\footnote{\label{fnRW}In the absence of drift
the length $X$ of the domain performs a symmetric random walk with the initial
condition $X(0)=L$. The closing time is equivalent to the first passage time in the origin, $X(t\ss{cl})=0$.
It is well known that for symmetric hopping its average value diverges, $\langle t\ss{cl}(L)\rangle=\infty$.
}
Any other choice of $L_2$ would be equally arbitrary: our choice is justified a priori by simplicity
and a posteriori by the comparison of our analysis with numerical results.

The COP case, Fig.~\ref{fig.models}(b), requires to consider two clusters of particles. This is because
two clusters can exchange matter until finally one of them disappears while 
a single cluster can never do that. 
For simplicity we assume that all domains (clusters of
particles and clusters of holes) have an initial length equal to $L$. 
If $T=0$ the configuration is frozen because the detachment of a particle requires energy.
If $T>0$, such process is permitted on the time scale $\tau\ss{det} \approx e^{4J(1)/T}$.
Once that a particle has detached from cluster ``1" it performs an asymmetric random walk 
ending its journey either reattaching to the same cluster or attaching to cluster ``2". 
In the former case the net outcome is null.
In the latter case there is a net exchange of mass ($1\to 2$) between the first and the second cluster and since 
the reversed process ($2\to 1$) may occur in first approximation with the same probability,\footnote{%
At the beginning of the process the two clusters are of equal length and the two possible exchanges of matter,
$1\to 2$ and $2\to 1$, are perfectly symmetric. In the course of time the different lengths between the two clusters
creates an asymmetry which will be neglected in our calculations.
}
there is a symmetric exchange of matter between the two clusters leading to a diffusion process for the length
$X(t)$, see Fig.~\ref{fig.models}(b). 

In Sec.~\ref{sec.COP} we will determine the average time $t\ss{cl}$ needed by a domain
of initial size $X(0)=L$ to either disappear ($X(t\ss{cl})=0$) 
or collect all the matter ($X(t\ss{cl})=2L$) due to the evaporation of the other domain.
The key ingredient to evaluate $t\ss{cl}$ will be the effective diffusivity of $X$, $D(L)$, which is inversely proportional
to the probability $p(L)$ that a monomer, detached from a cluster, attains the other one.

\section{Non conserved order parameter}
\label{sec.NCOP}

\subsection{Two domains approximation: analytical results}

Looking at Fig.~\ref{fig.models}(a),
as soon as $L_2 \ne L$ the domain walls feel a drift favoring the closure of the smallest domain.
As already said, we evaluate such drift for $L_2\to \infty$. The smaller domain has the time-dependent length $X(t)$, with $X(0)=L$.
If we define the integrated quantity
\be
I(x) \equiv \sum_{r=x}^\infty J(r) ,
\label{eq.I}
\ee
the process $X\to X+1$ requires the energy $(\Delta E)_+ = 4I(X)$,
while the process $X\to X-1$ releases the same energy.\footnote{More precisely, it is the process
$X+1 \to X$ to release the same energy, but for large $X$ we can neglect this difference.}
Therefore, using Eq.~(\ref{glauber}), the probabilities of such processes are
\be
p_\pm = \frac{1}{1 + e^{\pm 4\beta I(X)}},
\ee
and the drift is
\be
\delta(X) = p_+ - p_- = - \tanh (2\beta I(X)).
\label{exprdelta}
\ee

Next, we evaluate the closing time using a symmetric initial configuration, $L_2=L$
(see discussion related to footnote~\ref{fnRW}).
In terms of the random variable $X(t)$ whose evolution is controlled by
the probabilities $p_\pm$, this  amounts to have $X(0)=L$ and absorbing barriers in $X=0$ and $X=2L$.
Upon mapping the discrete, asymmetric random walk onto a convection-diffusion equation, and in
the approximation of constant drift (because of scaling it is assumed to depend on $X(0)$, not on $X(t)$),
the average exit time of the particle, given in~\cite{bookRedner}, is
\be
t(L) = \frac{L}{v} \tanh\left( \frac{vL}{D} \right). 
\label{tL}
\ee
This expression gives the correct limits $t(L)=L^2/D$ for vanishing drift and $t(L)=L/v$ for strong drift
($v\gg D/L$). According to the spirit of the above calculation and in view of Eq.~(\ref{exprdelta}),
the appropriate expression to be used for the drift is 
\be
v(L) = v_0\tanh (2\beta I(L)), 
\label{v}
\ee
where $v_0$ is a constant.
Therefore, Eqs.~(\ref{tL}) and (\ref{v}) give the closing time $t$ of a domain as a function of its initial size $L$.

It must be stressed that the simplified model describes the true situation 
with many domains if the next
boundary is at a distance where $J(r)$ is already very small.
This implies that the interaction with the subsequent boundaries can be
neglected, which is basically what the present approximation does.
Since the typical distance at which an interface is found is $L(t)$ we ask the condition $J(L)<e^{-M}$, 
where $M\gtrsim 1$, in order for the
approximation to be valid. This provides the lower limit $L_a$ for $L(t)$
above which the results of the model with few domains holds, with
\be
L_a\us{exp} = MR \qquad \mbox{and} \qquad
L_a\us{pow} = e^{M/(1+\sigma)}.
\label{ltyp}
\ee
Notice that $L_a$ depends linearly on $R$ for the exponential case
while it is weakly dependent on $\sigma$ for the algebraic case. 

Before studying Eqs.~(\ref{tL}) and (\ref{v}), let us write in the continuum approximation 
the explicit expression of the quantity $I(L)$ for an exponentially decaying interaction,
Eq.~(\ref{J_exp}), and for a power-law interaction, Eq.~(\ref{J_power}):
\bea
J\ss{exp}(r) =  e^{-r/R} &\qquad &  I\ss{exp}(L) \simeq  R e^{-L/R} \label{I_exp} \\
J\ss{pow}(r) =  \frac{1}{r^{1+\sigma}} &\qquad & I\ss{pow}(L) \simeq \frac{1}{\sigma} \frac{1}{L^\sigma} \; . \label{I_power}
\eea

In the following we demonstrate that Eqs.~(\ref{tL},\ref{v}) imply the existence of three dynamical regimes,
emerging as particular limits in which the arguments of the two hyperbolic tangents
appearing in Eqs.~(\ref{tL}) and (\ref{v}) are small or large. A summary of such regimes (plus others that are not captured by this analytical method) and of the characteristic crossover lengths between
them is provided in table~\ref{table}.

\vspace{.5cm}

{\it The ballistic regime ---}
In this regime $L(t)$ grows linearly with time, which means a constant $v$
(i.e., not depending on $L$).
This is possible if both the hyperbolic tangents in Eqs.~(\ref{tL}) and (\ref{v}) can be
approximated to one, i.e. if both arguments are very large,
\be
\frac{v_0 L}{D} \gg 1 \qquad \mbox{and} \qquad 2\beta I(L) \gg 1 .
\label{balineq}
\ee
The left inequality provides a lower limit for $L$, $L\gg (D/v_0)$, within
the approximation with few domains.
Notice however that this bound might be inadequate if $D/v_0$
happens to be smaller than the typical length $L_a$ above which  
the approximation holds true. 

On the other hand, the right inequality in~(\ref{balineq}),
considering the decreasing behavior of $I(L)$, provides un upper limit $L_b$ defined by the relation
$2\beta I(L_b) = 1$. More precisely, for the exponential and power-law models we find
\be
L_b\us{exp} = R \ln(2\beta  R) \qquad \mbox{and} \qquad
L_b\us{pow} = \left( \frac{2}{\sigma} \beta  \right)^{1/\sigma} ,
\label{ltypb}
\ee
which both diverge for $T\to 0$. In this limit, therefore, the ballistic dynamics becomes
the asymptotic one. For any finite $T$, instead, it is followed by the other regimes described below.

\vspace{.5cm}

{\it The slow regime ---}
This regime corresponds to a small $v$, such that the argument of the trascendental
function in Eq.~(\ref{v}) is small, and to a $t(L)$,
see Eq.~(\ref{tL}), which is given by the relation $t=L/v$. This means that we must have 
\be
2\beta I(L) \ll 1 \qquad \mbox{and} \qquad \frac{vL}{D} \gg 1 .
\label{eq.conNCOP}
\ee
The left inequality gives $L\gg L_b$ and  the resulting velocity is
\be
v(L) = 2 v_0 \beta I(L) .
\ee
Therefore, the right inequality of Eq.~(\ref{eq.conNCOP}) gives
\be
\frac{2 v_0}{D} L \beta I(L) \gg 1 ,
\label{in3}
\ee
which is consistent with the first of Eqs.~(\ref{eq.conNCOP}) and it allows us to define the length $L_c$ through the relation
\be
\frac{2 v_0}{D} L_c \beta I(L_c) = 1 .
\ee

For the exponential case we have
\be
\frac{2 v_0}{D} \beta  L\us{exp}_c R e^{-L\us{exp}_c/R} = 1,
\ee
which has the form $Ax e^{-x} = 1$, where $x=L\us{exp}_c/R$ and $A=\frac{2 v_0}{D} \beta  R^2$.
An approximate solution for $A\gg 1$ can be found as follows:
\be
\label{eqcorrlog}
x = \ln A + \ln x = \ln A + \ln\ln A + \ln\left( 1+ \frac{\ln x}{\ln A}\right) 
\simeq  \ln (A\ln A) .
\ee
So, we obtain
\be
L\us{exp}_c = R \ln\left[ \frac{2 v_0}{D} \beta  R^2 \ln\left( \frac{2 v_0}{D} \beta  R^2\right)\right] .
\ee

For the power-law case we have instead
\be
L\us{pow}_c = \left( \frac{2 v_0}{\sigma D} \beta \right)^{1/(\sigma -1)} 
= \left (\frac{v_0}{D}\right )^{1/(\sigma -1)} (L\us{pow}_b)^{\sigma/(\sigma -1)}.
\label{eq.Lcpow}
\ee
In the range $(L_b,L_c)$ the coarsening law is given by the relation
\be
t = \frac{L}{v(L)} = \frac{L}{2 v_0 \beta I(L)} ,
\ee
which implies a logarithmically slow dynamics for the exponential case,
\be
L(t) \sim R\ln t ,
\ee
and a non universal power-law growth for the power-law case,
\be
L(t) \simeq t^{1/(\sigma +1)} .
\ee

It is obvious that $L\us{pow}_c$, see Eq.~(\ref{eq.Lcpow}), is meaningless for $\sigma<1$. In fact, $L_a$ is a length of order one
and $L_b$ is a length which increases with decreasing the temperature or (for the exponential case) with increasing
$R$. These same properties apply to $L_c$ (and $L_c \gg L_b$) in the exponential case and in the power-law case
with $\sigma \ge 1$. Instead, if $\sigma < 1$ the inequality (\ref{in3}) writes
\be
L \gg \left( \frac{\sigma D}{2 v_0} \frac{1}{\beta } \right)^{1/(1-\sigma)} ,
\ee
which is automatically satisfied at low temperature, implying that the slow regime extends to infinity at any temperature $T<T_c$
for the power-law case with $\sigma <1$.
In all the other cases there will be a third regime, that we now discuss.

\vspace{.5cm}

{\it Diffusive regime ---} 
For $L \gg L_c$ it is $vL/D \ll 1$, so Eq.~(\ref{tL}) gives
\be
t = \frac{L}{v} \frac{vL}{D} = \frac{L^2}{D} ,
\ee
from which the usual growth law of the nn case,
\be
L(t) \simeq t^{1/2},
\ee
is recovered.

\begin{table}
  \centering
  \small
  \begin{tabular}{|c|c|c|c||c|c|c|c|c|c|}
\hline
& & & & & & & & & \\
&$\overset{\displaystyle \mbox{mean}}{\mbox{field}}$ &$L_{MF}$&
$\overset{\displaystyle \mbox{inter-}}{\mbox{mediate}}$ &$L_a$ & ballistic & $L_b$ & slow & $L_c$ & diffusive \\
& & & & & & & & & \\
\hline\hline
& & & & & & & & & \\
$J\ss{exp}$ & $e^t$ & $R$ & ??? &  $MR$ & $t$ & $R \ln(2\beta  R)$ & $\ln t$ &
$\overset{\displaystyle R \ln \{(2 v_0/D) \beta  R^2 \cdot }
  {\left . \ln\left[ (2 v_0/D) \beta  R^2\right]\right \}}$
& $t^{\frac{1}{2}}$\\
& & & & & & & & & \\
\hline
&  \multicolumn{3}{c||}{} & & & & & & \\
$\overset{\displaystyle J\ss{pow}}{\sigma > 1}$ & \multicolumn{3}{c||}{} &$e^{\frac{M}{1+\sigma}}$ &  $t$ & $\left( \frac{2}{\sigma} \beta  \right)^{\frac{1}{\sigma}}$ & $t^{\frac{1}{\sigma +1}}$ & 
$\left( \frac{2 v_0}{\sigma D} \beta \right)^{\frac{1}{\sigma -1}}$ & $t^{\frac{1}{2}}$ \\
&  \multicolumn{3}{c||}{} & & & & & & \\
\hline
&  \multicolumn{3}{c||}{} & & & & \multicolumn{3}{c|}{} \\
$\overset{\displaystyle J\ss{pow}}{\sigma \le 1}$ &  \multicolumn{3}{c||}{} &$e^{\frac{M}{1+\sigma}}$ & $t$ & $\left( \frac{2}{\sigma} \beta  \right)^{\frac{1}{\sigma}}$ & \multicolumn{3}{l|}{$t^{\frac{1}{\sigma +1}}$} \\
&  \multicolumn{3}{c||}{} & & & & \multicolumn{3}{c|}{} \\
\hline 
\end{tabular}
\caption{
Summary of the results for the NCOP class, for different models (see left column). Time ideally runs from left to right. 
  For each model we give on the top line the crossover scales $L_{MF},L_a,L_b,L_c$ where the various regimes start/end, and the corresponding regimes occuring between
  (or before or after) these lengths. In the lines below, the analytic expressions for the crossover lengths 
  and the behavior of $L(t)$ in the corresponding regimes is reported (for the intermediate regime, see the question marks, 
  $L(t)$ is not analytically known).
  The quantities to the right of the double vertical
  line are those predicted by the approximation with few domains.
  Some regimes, and the corresponding crossover lengths, do not exist for certain models. 
  For example, for the power model with $\sigma \le 1$ the length $L_c$ 
is not defined, the diffusive stage does not exists and the slow regime is the asymptotic one.}
\label{table}
\end{table}

We summarize all results for NCOP coarsening in Table~\ref{table}.
The regimes described by the approximation with few domains are those occurring
for $L(t)>L_a$, namely those to the right of the double vertical line in the
Table. The early regimes with $L(t)<L_a$, which are observed only for
$J\ss{exp}$, will be discussed later.

\subsection{Simulations: interactions decaying exponentially}

In Fig.~\ref{fig_Lt_exp_NCOP} we plot the simulation results for $J\ss{exp}(r)$ with varying $R$. 
In this figure we report results of simulations obtained with both methods 
of Sec.~\ref{simuldes}.
Clearly, with the  method S1 a much smaller time range can be investigated.
When using method S2, bulk flips are forbidden, and for any $R$ we searched for a value of $n_K$ representing a good
compromise between the accuracy of the simulations and their efficiency (for the largest values of
$R$, $n_K$ is of order $2\cdot 10^2$).
For small values of $R$ the two kind of calculations agree rather well at any time. This can be clearly seen
in the case with $R=5$. The two methods also agree, for any $R$, at sufficiently long times, as
expected. Besides that, for any $R$ they also agree in a very early regime where a fast exponential 
growth of $L(t)$ takes place. This regime is not captured by our previous analytical arguments
because, recalling the discussion below Eq.~(\ref{balineq}), it occurs
when $L(t)$ is still too small, $L(t)<L_a$, for the model with a single domain
to be adequate. Such regime, which is of a mean field character, will be worked out analytically in Sec.~\ref{stmfr}. 
On the other hand, for large
$R$, simulations S1 and S2 disagree in an intermediate time interval after the mean field regime.
For instance, for $R=10^2$ the curves obtained with methods S1 and S2 are quite different in the time
range between $t\simeq 10$ and $t\simeq 10^3$. We remind that this discrepancy is partly due to the
limited number $n_K$ of kinks considered but, especially, to the fact that flips in the bulk cannot be
neglected in this time domain. 

Regarding the behavior of $L(t)$, the first observation is the impressive difference, particularly
for large $R$, with respect to the behavior of the nn model.
Looking at the curves obtained with the simulation method S2 the three regimes discussed above, ballistic ($L\sim t$), slow ($L\sim \ln t$), and diffusive ($L\sim t^{1/2}$) are clearly visible, and the crossovers between them occur
at values of $L(t)$, $L\us{exp}_a,L\us{exp}_b,L\us{exp}_c$, which are consistent
with their estimations given in Table~\ref{table} or equivalently in
Eqs.~(\ref{ltyp},\ref{ltypb},\ref{eq.Lcpow}). 
These estimations say that,
with increasing $R$ the ballistic regime ends later, the slow kinetics lasts
longer, and the diffusive regime starts later.
In the exact simulations performed with method S1 the ballistic regime
is only barely observed for $R=10^2$, since the range of times where it shows up is small. 
It should be more clearly observed for larger $R$ but the numerical effort to go to (reasonably) larger
values of $R$ turns out to be much beyond the scope of this paper.   
  
It is interesting to display the analytical curves for $L(t)$, obtained
solving numerically the two coupled equations (\ref{tL}) and (\ref{v}), see Fig.~\ref{th_nc}. 
Here the three different regimes are very clearly observed and the figure has the merit of stressing
that the behavior of $L(t)$ in the ballistic and the diffusive regimes does not depend on $R$ (while, clearly, the duration of such regimes, namely the quantities
$L_a\us{exp},L_b\us{exp},L_c\us{exp}$, do depend on $R$).
This is true but less evident for simulations as well, see Fig.~\ref{fig_Lt_exp_NCOP}.
To better analyze this point one can have a look to the lower inset of Fig.~\ref{fig_Lt_exp_NCOP}
where the same data of the main panel are plotted on rescaled axes.
Although the flat plateau occurring when the fast initial growth of $L(t)$ is over is a spurious
effect due to use of method S2 for simulating the system (indeed, as it can be seen in the upper
panel this plateau is washed out in the exact simulations) it is instructive to study its
location.
Its height lies around $L(t)/R\simeq 10$, and after that
the ballistic regime starts.
This shows that, as already discussed, the convective regime starts at
$L(t)=L_a\sim MR$, with $M\simeq 10$. Since the ballistic regime is independent
of $R$ this implies that it starts at a time of order $R$, and indeed in the
figure it is seen that it begins at $t/R\simeq 1$.

\bfi
\begin{center}
  \includegraphics*[width=0.99\columnwidth]{fig_Lt_exp_NCOP.eps}
\end{center}
\caption{
{\bf (NCOP exp)}
In the main panel the domain size $L(t)$ for a quench from $T_i=\infty$ to $T=10^{-2}$, 
for a system of size $N=10^7$, is shown on a log-log plot. Continuous lines with symbols are
obtained by means of simulations using method S2.
Different curves correspond to the nn interaction and to the interaction $J\ss{exp}(r)$
with different values of $R$, as indicated. For some values of $R$ ($R=5,10^2,10^4$) we plot
also the curve obtained with the simulation method S1, see dotted lines. 
In the simulation method S1 we set ${\cal M}=100$ and we have verified that no significant change is observed for larger ${\cal M}$ values. 
The dashed green and violet lines are the algebraic forms $t^{1/2}$ and $t$, respectively. 
The curve drawn with turquoise + symbols at short times represents the exponential growth in the 
mean-field regime, obtained analytically in Eq.~(\ref{eq.MF}).
In the upper inset the data for $R=10^4$ are plotted on a log-linear scale to appreciate the logarithmic behavior.
The lower inset shows the same data of the main panel (only simulations with method S2 for $R=10^2,10^3,10^4$), 
but plotting the rescaled quantity $L(t)/R$ against $t/R$.
}
\label{fig_Lt_exp_NCOP}
\efi

\bfi
\includegraphics*[width=0.95\columnwidth]{th_nc.eps}
\caption{
{\bf (NCOP exp)}
Plot of $L(t)$, obtained from reversing the function $t(L)$ as given by Eqs.~(\ref{tL}) and (\ref{v}), on a double logarithmic scale.
Upper left inset: we plot the same data in the short time regime with linear scales to show the ballistic regime.
Lower right inset: the same data are plotted on a log-linear scale to
show the logarithmic regime.}
\label{th_nc}
\efi

\subsubsection{Short time mean-field regime} \label{stmfr}

We want now to discuss the early regime, characterized by a fast (exponential) growth of order.
Since this occurs for values of $L(t)$ smaller than $L_a$,
the approximation with few domains fails. 
Luckily, in the limit $L \ll R$ a mean field
approximation holds. We are therefore going to discuss such an approximation.

If any spin interacts with other spins within a distance $R\gg 1$ 
the magnetization density $m$ increases in time following the equation~\cite{EPL}
$dm/dt= [-m + \tanh(2\beta R m)]$.
The initial value of $m$ is the result of the imbalance between positive and negative spins on a scale 
of order $R$. Because of the central limit theorem,
$m(0) \simeq 1/\sqrt{R}$ (we assume $m(0)>0$) so that the minimal 
value of the argument of $\tanh$ is of order $2\beta \sqrt{R}\gg 1$.
Therefore, the approximation $\tanh(2\beta R m) \simeq 1$ is correct and 
we can write $dm/dt = (1-m)$, whose solution is $m(t) = 1 - [1-m(0)]\exp(-t)$. 

Because of the random character of the initial configuration, and due to the fact that
the mean field dynamics does not introduce any correlation among spins,
the configuration corresponding to a magnetization density $m$
can be obtained by choosing the $i-$th spin as $s_i=+1$ with probability $p_+=(1+m)/2$ and as $s_i=-1$ with probability $p_-=(1-m)/2$. 
In doing that, the probability of having $\ell_+ $ consecutive positively aligned spins
is $P(\ell_+)=p_-p_+^{\ell_+-1}$ and the average length of such domain is
$\bar\ell_+=1/p_-$. Analogously, for negative spins one finds
$\bar\ell_-=1/p_+$.
Therefore the average domain size is
$L=\frac{1}{2}(\bar\ell_+ +\bar\ell_-)=1/(2p_- p_+)$.
Using the explicit expression for $p_{\pm}$ in terms of the magnetization, we obtain
\be
L(t) = \frac{2}{1-m^2(t)} = \frac{2}{1+m(t)}\cdot \frac{e^t}{1-m(0)} . 
\label{eq.MF}
\ee
Since the quantity $1+m(t)$ is very weakly dependent on $t$, it varies from $(1+m(0))$ to $2$, this equation clearly
shows the exponential growth of $L(t)$ in the mean field regime, $L(t)\simeq e^t$.
In Fig.~\ref{fig_Lt_exp_NCOP} we can check that simulation results well reproduce the
prediction of Eq.~(\ref{eq.MF}) (compare the dotted magenta line with the + turquoise symbols) at short times.

The exponential growth ends when $L(t)$ is of order $R$, which happens at a time of order $\ln R$. 
Recalling that the ballistic regime starts at $t/R\simeq 1$
and for $L(t)=L_a=MR$, with $M\simeq 10$, 
there is a time lag
of order $R-\ln R\simeq R$ in which $L(t)$ must fill the gap from the
end of the mean-field regime with $L\simeq R$ to the beginning of the
ballistic one with $L\simeq MR$. In this {\it intermediate} regime both the mean-field
approximation and the one with few domains fail, because one has neither
many boundaries nor a single boundary within the interaction distance. Also, this is a regime
where simulations of kind S2 fail (basically for the same reason). 
We could not devise any scheme to derive quantitative informations in this intermediate time range.

\subsection{Simulations: interactions decaying algebraically}

Before starting our discussion of the case with algebraic interactions let us comment on the fact that,
at variance with the exponential case, now the use of method S2 for the simulations can alway be made reliable, at any time, by 
tuning $n_K$ appropriately.
Indeed, as shown in Fig.~\ref{fig_Lt_pw_NCOP}, the initial mean-field regime is never present in this case and the approximated method S2 always correspond to neglect terms of the order of $\left (n_k L(t)\right )^{-1-\sigma}$ which become sub-leading for a sufficienly large $n_k$. In particular,  
 we have compared S2 simulations with the exact ones for several choices of the parameters
$\sigma$ and $T$ and we found excellent agreement in the whole time domain using values 
at most equal to $n_K=200$. In the following, therefore,
we will present always data obtained with method S2.

Let us remind that for $J(r)=J\ss{pow}(r)$ our theory predicts a ballistic regime ($L\sim t$), then a slow regime
($L\sim t^{1/(1+\sigma)}$) which is asymptotic for $\sigma \le 1$ or is followed by a final diffusive regime ($L\sim t^{1/2}$) if $\sigma>1$.
In Fig.~\ref{fig_Lt_pw_NCOP} we plot the NCOP results for different values of $\sigma$
and for the nn case.
Starting from the smallest value of $\sigma $, namely $\sigma =0.5$,
one observes that, after a short transient for $t\lesssim 10$, or equivalently
for $L(t)\lesssim L\us{pow}_a \simeq 10^2$, the ballistic regime is entered which extends
up to the longest simulated time $t_{max}$. For this choice of the parameters, indeed, the crossover
to the slow regime is delayed after $t_{max}$, as we will prove in a while.
The curves for increasing values of $\sigma$ superimpose on the $\sigma =0.5$ curve
up to a certain value $L\us{pow}_b$ of $L$ which gets smaller the larger $\sigma $ is,
according to Eq.~(\ref{ltyp}). For $\sigma =3$, $L\us{pow}_b$ is so small that
the ballistic regime is not even observed. The late stage diffusive regime that is
expected for $\sigma >1$ is not observed for $\sigma =1.5$ because it sets in
at $L(t)=L\us{pow}_c$ but this quantity, according to Eq.~(\ref{eq.Lcpow}),
diverges for $\sigma \to 1$ and hence cannot be reached in our simulations.
On the other hand, the diffusive regime is very neatly observed for $\sigma =3$ and
is incepient for $\sigma =2$ at late times. Notice also in this case, as for
the exponential coupling, the profound difference with respect to the nn case.

In order to show the transition to the slow regime also for the case with
$\sigma =0.5$ we show, in Fig.~\ref{fig_Lt_sigmas_NCOP}, $L(t)$
for quenches to different temperatures. Indeed, 
the slow regime starts at the length scale
$L_b^{pow}$ of Eq.~(\ref{ltypb})
which, although typically very large when $\sigma<1$, can be reduced by
increasing the temperature. This figure shows the crossover to the slow regime
as $T$ is raised, as expected.

\bfi
 \centering
 \rotatebox{0}{\resizebox{.85\textwidth}{!}{\includegraphics{fig_Lt_pw_NCOP.eps}}}
\caption{
{\bf (NCOP pow)}
  $L(t)$ for NCOP quenched from $T_i=\infty$ to $T=10^{-3}$ on a double-logarithmic scale.
  Different symbols and colors
correspond to different values of
 $\sigma$ and to the nn case(see legend).  The dashed orange line
 is the $t^{1/2}$ law and the dashed green one is the ballistic behavior.  
 The color dotted lines (below the data curves) are the power-laws
 $t^{1/(\sigma+1)}$ of the slow regime for each $\sigma$ value.
}
\label{fig_Lt_pw_NCOP}
\efi

\bfi
 \centering
 \rotatebox{0}{\resizebox{.85\textwidth}{!}{\includegraphics{fig_Lt_sigmas_NCOP.eps}}}
\caption{
{\bf (NCOP pow)}
$L(t)$ for NCOP quenched from $T_i=\infty$ to different final $T$ for $\sigma=0.5$.  The green dashed line 
  indicate the linear, ballistic regime $L(t)\sim t$.
  The magenta dashed line is the growth
$L\sim t^{1/(\sigma+1)}=t^{2/3}$ in the slow regime.
  The system size is $N=8\times 10^8$.}
\label{fig_Lt_sigmas_NCOP}
\efi

\section{Conserved order parameter}
\label{sec.COP}

\subsection{Two clusters approximation: analytical results}

If the order parameter is conserved a domain disappears because
all its particles evaporate and attach to neighbouring clusters. There are now two relevant random walk processes, 
see Fig.~\ref{fig.models}(b):
(i)~each cluster emits and absorbs particles, so that its length $X(t)$ is an integer positive random variable which can increase or decrease;
(ii)~each emitted particle performs a random walk between two clusters and $X(t)$ actually increases or decreases
only if an emitted particle is absorbed by a different cluster.

The minimal model involves two clusters of particles in a ring geometry and, for simplicity, we are
going to consider two clusters of initial length $L$, separated by a distance $L$. In spin language
the initial configuration is composed by four domains of equal length. 
In this way the process is (at least initially) perfectly symmetric and we can limit to determine the probability $p(L)$ 
that a particle emitted by cluster 1 is absorbed by cluster 2.
Once we know $p(L)$ the resulting closing time of one of the two domains is simply
\be
t(L) = t_0 L^2 /p(L),
\ee
where $t_0 =e^{4\beta J(1)}$ is the characteristic time of particle emission and
$t_0/p(L)$ is the typical time because a cluster varies its size by one.

In the nn model a particle between two neighbouring clusters diffuses freely and it is straightforward to
derive that $p(L) = 1/L$, as argued below~\cite{Nato_Rodi}. 
For symmetry reasons $p(L)=\frac{1}{2} p(L/2)$ because once the particle has attained 
an equal distance to both clusters the probability to attach to the right cluster is equal
to the probability to attach to the left cluster. Therefore, $p(L)=\frac{a}{L}$ and
since $p(2)=\frac{1}{2}$, we obtain $a=1$.
If the range of $J(r)$ is not limited to nn, 
the random walk of the particle emitted by a cluster is not symmetric. 
Indeed, when a monomer detaches from a cluster it feels the attraction from the same cluster, 
a fact that strongly reduces the probability $p(L)$ with respect to the nn case.
As we will see, this produces an overall slow down of the kinetics upon increasing the interaction range.

We are now going to determine the drift $\delta(x)$ felt by a particle at distance $x$ from the closest domain,
see Fig.~\ref{fig.models}(b).
The energy $E(x)$ associated to the particle can be evinced from Eq.~(\ref{energygaslat}) by
singling out in all the sums the term with $i=i_x$, $i_x$ being the site where the monomer
is. Neglecting irrelevant additive constants which are independent of $L$ one has
\be
E(x) = - 4\left( \sum_{r=x}^{x+L} J(r) + \sum_{r=L-x}^{2L-x} J(r) \right) , 
\ee
which is a function defined in the interval $(0,L)$ and symmetric with respect to the midpoint, $x=L/2$, where it has a maximum.
It is obvious that the drift is negative for $x<L/2$ and it is positive for $x>L/2$. 
For symmetry reasons it is always true that $p(L)=\frac{1}{2}\tilde p(L/2)$ where $\tilde p(L/2)$ is the probability to attain
the midpoint $x=L/2$ and, in order to evaluate this probability, we only need the negative drift $\delta(x)$ for $x<L/2$.
In this interval we can forget the effect of the farthest domain and we can assume a diverging length of the nearest domain,
so that for $x<L/2$ we can approximate $E(x)$ as follows,
\be
E(x) \simeq -4 \sum_{r=x}^\infty J(r) = -4 I(x) .
\ee

The probability $p_+(x)$ to hop from $x$ to $x+1$ is 
\be
p_+(x) = \frac{1}{1+ e^{\beta(E(x+1)-E(x))}} = \frac{1}{1+ e^{4\beta J(x)}}
\ee
while the probability to hop from $x$ to $x-1$ is
\be
p_-(x) = \frac{1}{1+ e^{-4\beta J(x)}} ,
\ee
so that 
\be
\delta(x) = p_+(x) - p_-(x) = - \tanh(2\beta J(x)).
\label{eq.deltaCOP}
\ee

We now must determine $\tilde p(L/2)$ from the knowledge of $\delta(x)$. Let us consider the following first-passage problem:
a particle diffuses anisotropically on the integer sites of the interval $[0,N]$ and we wonder what is the probability
$W_N(x)$ that a particle currently at $x$ reaches the site $N$ before reaching the site $0$. We can write
\be
W_N(x) = p_+(x) W_N(x+1) + p_-(x) W_N(x-1) .
\ee
In a continuum approximation, writing $W_N(x\pm 1)\simeq W_N(x)\pm W_N'(x)+(1/2)W_N''(x)$, using the
relation $p_+ +p_-=1$ and Eq.~(\ref{eq.deltaCOP}),
the above equation reads $W_N''(x) = -2\delta(x) W_N'(x)$, which should be supplemented with 
the boundary conditions $W_N(0)=0$ and $W_N(N)=1$.
The equation can be integrated twice, giving
\be
W_N(x) = \frac{ \int_0^x dy\, e^{-2\int_0^y ds \delta(s)} }{ \int_0^N dy\,e^{-2\int_0^y ds \delta(s)} }.
\ee
Observing that $\tilde p(L/2)=W_N(x)$ for $x=1$ (the site where a detached particle starts to diffuse) and $N=L/2$,
we can write
\be
\tilde p(L/2) = \frac{ \int_0^1 dy\, e^{2\int_0^y ds |\delta(s)|} }{ \int_0^{L/2} dy\, e^{2\int_0^y ds |\delta(s)|} } \equiv
\frac{\cal A}{{\cal B}(L)} ,
\ee
 with $|\delta(s)| =\tanh(2\beta J(s))$. With the knowledge of ${\cal A}$ and ${\cal B}$, we have the following relation to
determine the coarsening law,
\be
t = \frac{2 t_0}{\cal A} L^2 {\cal B}(L) .
\label{eq.COP}
\ee

Let us now evaluate the quantity ${\cal B}(L)$.
  Starting with the expression for $|\delta(s)|$, the argument of the hyperbolic tangent, $2\beta J(s)$,
varies from $2\beta J(s) \gg 1$ at small $s$ to $0$  for diverging $s$, so there is a crossover between two regimes. If $s^*$ is defined
by the relation $2\beta J(s^*)=1$, we have
\be
|\delta(s)| \simeq \left\{ 
\begin{array}{lcc}
1 & & s \ll s^* \\
2\beta J(s) & & s \gg s^*
\end{array}
\right. ,
\label{moddelta}
\ee
with a sharp transition between the two regimes. We can therefore approximate the integral appearing in the exponent as follows,
\be
C(y) \equiv \int_0^y ds |\delta(s)| \simeq \left\{
\begin{array}{lcr}
y, & & y < s^* \\
s^* + 2\beta [ I(s^*) - I(y) ], &\quad & y > s^*
\end{array}
\right. .
\ee

We finally obtain
\be
{\cal B}(L) = \int_0^{L/2} dy \,e^{2 C(y)} \simeq \left\{
\begin{array}{lcr}
\int_0^{L/2} dy\; e^{2y} = \frac{1}{2}(e^L -1) \simeq \frac{e^L}{2}, && \frac{L}{2} < s^* \\
& & \\
\frac{e^{2s^*}}{2} + e^{2s^*} \int_{s^*}^{L/2} dy\; e^{4\beta [ I(s^*) - I(y)]}, &\qquad & \frac{L}{2} > s^*
\end{array}
\right. .
\ee

As for the numerator, 
\be
{\cal A} = \int_0^1 dy\; e^{2 C(y)} \simeq \int_0^1 dy\; e^{2y} = \frac{1}{2}(e^2 -1) \equiv c_0 \simeq 3.1945.
\ee

We can now find the limiting behaviors for small $L$ and diverging $L$. For $L< 2s^* = L_s$ we obtain
\be
t = \frac{t_0}{c_0} L^2 e^L, \qquad L \ll L_s
\ee
which gives a logarithmically slow coarsening, $L(t) \simeq \ln t$.

For the opposite case, $L>L_s$, we can observe that $(I(s^*) - I(y))$ is an increasing function from $y=0$ (for $s=s^*$)
to the positive, constant value $I(s^*)$ for $y\to\infty$. So, we expect the leading term for diverging $L$ 
to be
\be
{\cal B}(L) \simeq e^{L_s} e^{4\beta I(L_s/2)} \frac{L}{2} ,
\ee
and
\be
t \simeq \frac{t_0}{c_0} e^{L_s} e^{4\beta I(L_s/2)} L^3 , \qquad L\to \infty
\label{eq.asCOP}
\ee
thus obtaining $L \approx t^{1/3}$ asymptotically. Notice that, quite interestingly,
the logarithmic and the
diffusive regime do not make explicit reference to the form of the interaction.

However, we must observe that the transition of $(I(s^*) - I(y))$ from zero to $I(s^*)$ depends on the 
explicit form of the coupling: in the exponential case, $J\ss{exp}(r)$ rapidly decays to zero and such transition is sharp;
in the power law case, $J\ss{pow}(r)$ does not decay rapidly, the transition is not sharp and a third intermediate regime
exists, as we are going to argue.

For $J(r)=1/r^{1+\sigma}$, $I(y)=1/(\sigma y^\sigma)$ and for $L>L_s$ it is
\bea
{\cal B}(L) &=& \frac{e^{2s^*}}{2} + e^{2s^*} e^{\frac{4\beta}{\sigma}\frac{1}{(s^*)^\sigma}} 
\int_{s^*}^{L/2} dy e^{-\frac{4\beta}{\sigma}\frac{1}{y^\sigma}} \\
&=&
\frac{e^{2s^*}}{2} + e^{2s^*} e^{\frac{4\beta}{\sigma}\frac{1}{(s^*)^\sigma}}
\frac{1}{\sigma} \left( \frac{4\beta}{\sigma} \right)^{1/\sigma}
\left[ 
\Gamma\left( -\frac{1}{\sigma} , \frac{4\beta}{\sigma} \left( \frac{2}{L} \right)^{\sigma} \right) -
\Gamma\left( -\frac{1}{\sigma} , \frac{4\beta}{\sigma} \left( \frac{1}{s^*} \right)^{\sigma} \right) 
\right] ,
\label{eq.inGamma}
\eea
where $\Gamma(\alpha,x)$ is the upper incomplete Gamma function.

We define the length $L\ss{cr}$ 
such that the argument $x$ of the Gamma function on the left between square brackets, Eq.~(\ref{eq.inGamma}), is equal to one.
We find 
\be
L\ss{cr} = 2^{1+(2/\sigma)} (\beta/\sigma)^{1/\sigma},
\label{lcr}
\ee
 and 
$L\ss{cr}/L_s=\frac{2^{(2+\sigma)/[\sigma (1+\sigma)]}}{\sigma^{1/\sigma}} 
\beta^{1/(\sigma(1+\sigma))} \gg 1$.

For $L\gg L\ss{cr}$ the incomplete Gamma function on the right between square brackets, Eq.~(\ref{eq.inGamma}),
can be neglected with respect to the left $\Gamma$ and using the asymptotic
expansion for $x\ll 1$ we obtain Eq.~(\ref{eq.asCOP}).
For $L_s \ll L \ll L\ss{cr}$ we can use the expansion valid for $x\gg 1$,
$\Gamma(-1/\sigma,x) \simeq x^{-1-(1/\sigma)} e^{-x}$, and we obtain
\be
t \simeq \frac{t_0}{c_0 2^{\sigma+2}} \frac{e^{L_s}}{\beta} L^{\sigma +3} 
\exp\left[
\frac{\beta 2^{\sigma+2}}{\sigma}\left( \frac{1}{(L_s)^\sigma} - \frac{1}{L^\sigma}\right)\right] 
\qquad (L_s \ll L \ll L\ss{cr}).
\label{eq.mipoCOP}
\ee
In the limiting case $\sigma\to 0$ we have $L_s=4\beta$ and $L\ss{cr} \to\infty$,
so that the intermediate regime is the asymptotic regime.
In the same limit we find
\be
t \simeq \frac{t_0}{c_0} \left(\frac{e}{4\beta}\right)^{4\beta}
\frac{L^{4\beta +3}}{4\beta} 
\qquad (\sigma\to 0^+).
\label{eq.sigma0}
\ee
A summary of the regimes and of the crossover lengths separating them is given in Table~\ref{tablecop}.

\begin{table}
  \centering
  \small
  \begin{tabular}{|c|c|c|c|c|c|}
\hline
& & & & & \\
& slow & $L_s$ & intermediate & $L_{cr}$ & diffusive \\
& & & & & \\
\hline\hline
& & & \multicolumn{2}{c|}{} & \\
$J\ss{exp}$ & $\ln t$ & $2R\ln(2\beta )$ & \multicolumn{2}{c|}{} & $t^{\frac{1}{3}}$\\
& & & \multicolumn{2}{c|}{} & \\
\hline
& & & & & \\
$\overset{\displaystyle J\ss{pow}}{\sigma > 0}$ & $\ln t$ & 
$2(2\beta )^{\frac{1}{1+\sigma}}$ & Eq.~(\ref{eq.mipoCOP}) &$ 2^{1+(2/\sigma)} (\beta/\sigma)^{1/\sigma}$& $t^{\frac{1}{3}}$ \\
& & & & & \\
\hline
& & & \multicolumn{3}{c|}{} \\
$\overset{\displaystyle J\ss{pow}}{\sigma \to 0^+}$ & $\ln t$ & 
$2(2\beta )^{\frac{1}{1+\sigma}}$ & \multicolumn{3}{l|}{$t^{\frac{1}{4\beta +3}}$}  \\
& & & \multicolumn{3}{c|}{} \\
\hline 
\end{tabular}
\caption{Summary of the results for the COP class. Time ideally runs from left to right. 
 On the top line the scales $L_s,L_{cr}$ where the various regimes start/end, and the corresponding regimes occuring between
  (or before or after) these lengths are indicated. In the lines below, the analytic expressions for the crossover lengths 
  and the behavior of $L(t)$ in the corresponding regimes is reported, for the different forms of $J(r)$.  Some regimes, and the corresponding crossover
  lengths, do not exist for certain models. 
 For example, for the exponential model there is no intermediate regime because $L_s$ and $L_{cr}$ are not distinct.}
\label{tablecop}
\end{table}

\subsection{Simulations: interactions decaying exponentially}

For an exponential coupling, $J\ss{exp}(r)$, our theory predicts a logarithmic coarsening, for $L(t)<L_s=2s^*$, followed by the
asymptotic, power-law regime, $L\sim t^{1/3}$. Using the definition of
$s^*$ given above Eq.~(\ref{moddelta}) we have
\be
L_s\us{exp}=2R\ln(2\beta).
\label{lsubs}
\ee
These two regimes and the sharp crossover between them
are very neatly observed by plotting the theoretical formulas in
Fig.~\ref{fig_cop_anal}. Notice also that the crossover length scales
as $R$, as expected according to Eq.~(\ref{lsubs}).

\bfi
\includegraphics*[width=0.95\columnwidth]{fig_cop_anal.eps}
\caption{ {\bf (COP exp)}
  $L(t)$ for COP quenched to $T =0.1$ according to the analytical
  prediction based on the model with few domains, Eq.~(\ref{eq.COP}),
  with $J(r)=J\ss{exp}(r)$ and various values of $R$, see key.
  The dashed violet line is the asymptotic diffusive behavior $L(t)\sim t^{1/3}$.
  In the main figure the data are plotted on a double logarithmic
  scale, and the heavy circles correspond to the crossover lengths of Eq.~(\ref{lsubs}) (for 
  $R=10$ it is beyond the largest time in the plot). The inset shows the same data but using a logarithmic-linear scale
  in order to show the initial logarithmic regime.}
\label{fig_cop_anal}
\efi

Let us now discuss the results of our simulations performed according to the method S2 discussed 
in Sec.~\ref{simuldes}. We have checked that with COP this kind of simulations provide
reliable results basically at any time, provided that $n_K$ is chosen appropriately 
(mostly, we used $n_K=2\cdot 10^2$). This is true for both kinds of interactions $J(r)$ considered.
Despite the speed-up provided by simulations with a reduced number of kinks, 
calculations with COP are quite time demanding and, therefore, it is difficult
to push $R$ to rather large values. Nevertheless, even if  data display
the transition between the two regimes less clearly than the solution of
the model with few domains, there is still a rather clear
evidence of a short-time logarithmic coarsening and of the asymptotic $t^{1/3}$ 
power law following it, see Fig.~\ref{fig_Lt_exp_COP}. Also, the crossover between
them is delayed by increasing $R$, as expected.
Notice that in our analytical approach the logarithmic regime turns out to be 
independent of $R$. Indeed curves for different $R$ superimpose in Fig.~\ref{fig_cop_anal}.
This is not observed in the numerical simulations of Fig.~\ref{fig_Lt_exp_COP}.
This could be possibly due to an offset caused by a very early regime which is not captured by
our analytical techniques, or/and to the impossibility to reach sufficiently large values
of $R$ in simulations.

\bfi
\includegraphics*[width=0.95\columnwidth]{fig_Lt_exp_COP.eps}
\caption{ {\bf (COP exp)}
$L(t)$ for a quench from $T_i=\infty$ to $T=0.4$ for a system size $N=10^7$ and different values of $R$, as detailed in the key. 
  The plot in the main figure is on a double-logarithmic scale. The dashed
  green line is the the asymptotic diffusive behavior $L(t)\sim t^{1/3}$.
The horizontal dotted line corresponds to $L\ss{nn}(T=0)\simeq 4.135$, the asymptotic domain size for the nn model
at $T=0$.
  In the inset the same data (only for $R=2$) are plotted with log-linear scales
so as to show the initial, logarithmically slow coarsening.
}
\label{fig_Lt_exp_COP}
\efi

\subsection{Simulations: interactions decaying algebraically}

In the power law case we still expect the logarithmic regime and the asymptotic, $t^{1/3}$, regime, but
in addition we should observe an intermediate regime, as given by Eq.~(\ref{eq.mipoCOP}), for
$L_s<L(t)<L_{cr}$ , with (see Table~\ref{tablecop}):
\be
L_s\us{pow}=2(2\beta )^{\frac{1}{1+\sigma}} \mbox{~~~and~~~}
L_{cr} = 2^{1+(2/\sigma)} (\beta/\sigma)^{1/\sigma} .
\label{lsubspow}
\ee
These analytical predictions are corroborated by the explicit numerical solution of Eq.~(\ref{eq.COP}),
see Fig.~\ref{cop_formula_Paolo}. First of all the asymptotic power law expected for $\sigma\to 0^+$ is 
very well observed. For the largest values of $\sigma $ we have considered, namely $\sigma =1$ and $\sigma =2$,
after a relatively short preasymptotic regime, $L(t)$ attains the asymptotic diffusive behavior
$L(t)\sim t^{1/3}$. As $\sigma $ is lowered the preasymptotic stage increases, because both
$L_s$ and $L_{cr}$ increase, and the asymptotic stage is pushed beyond the scale of times
presented in the figure. In the rightmost inset of Fig.~\ref{cop_formula_Paolo} we show the early
time behavior of the preasymptotic stage with linear-log scales, in order to detect the logarithmic law.
Although the growth is definitely slower than an algebraic one, the resulting plot is not fully linear, not even for the smaller values of $\sigma$, because there is a tiny upward curvature. This is perhaps due
to the fact that the crossover to the next stages is broad.
In addition, the leftmost 
inset shows that, after this slow regime, an intermediate regime where Eq.~(\ref{eq.mipoCOP}) holds
is observed. Actually, according to this equation, by defining $\Lambda^{-1}=\frac{\beta 2^{\sigma +2}}{\sigma}L^{-\sigma}$, 
this quantity should behave as
$\Lambda ^{-1}=\mbox{const}-\tau$, where
$\tau=\ln \left (\frac{t}{L^{\sigma+3}}\right )$. Indeed, a linear relation between $\Lambda ^{-1} $ and $\tau$ is 
observed in the inset of Fig.~\ref{cop_formula_Paolo} after a certain $\tau$.

\bfi
 \centering
\includegraphics[width=0.8\textwidth,clip=true]{cop_formula_Paolo.eps}
\caption{ {\bf (COP pow)}
$L(t)$ for a quench to $T =0.1$ according to the model with few domains, Eq.~(\ref{eq.COP}),
for $J(r)=J\ss{pow}(r)$ and with different values of $\sigma $, see key.
In the main part of the figure, data are plotted on a log-log scale.
The green dashed line is the behavior $t^{\frac{1}{4\beta +3}}$ expected for
$\sigma\to 0^+$. The violet dashed line is the diffusive behavior $t^{1/3}$.
In the rightmost inset the same data are plotted on a log-linear scale, to show
the logarithmic regime. In the leftmost inset the quantity
$\Lambda^{-1}=\frac{\beta 2^{\sigma +2}}{\sigma}L^{-\sigma}$ is plotted against
$\tau=\ln \left (\frac{t}{L^{\sigma+3}}\right )$. According to
Eq.~(\ref{eq.mipoCOP}) one should have the intermediate regime $\Lambda=\mbox{cost} -\tau$,
which is indeed well observed.}
\label{cop_formula_Paolo}
\efi

When we pass to simulations, the power law COP model is more elusive.
Firstly, it is not possible to clearly show both logarithmic and
$t^{1/3}$ regimes for the same parameters, because simulations cannot access the
whole range of time that would be needed. Indeed, if $T$ is sufficiently high, given the form of 
$L_s$ and $L_{cr}$, Eq.~(\ref{lsubspow}), one is able to enter the asymptotic
diffusive regime but the preasymptotic ones are too compressed to be observable.
On the contrary, lowering $T$ one is able to see the preasymptotic regime (at least the
slow logarithmic one, see below), particularly for small $\sigma$, but the asymptotic one
is so delayed to be unreachable.
For this reason we present in the following two figures where we change $\sigma $ and 
$T$ separately in order to observe the early as well as the late regimes.
Specifically, in Fig.~\ref{fig_Lt_pw_COP} we show a quench to a temperature $T=0.4$ which is
large enough to observe the asymptotic regime. We see that $L(t)\sim t^{1/3}$ is observed
at late times for $\sigma =2$ and $\sigma =3$. For smaller values of $\sigma $ the crossover
to the asymptotic stage is at most incipient (for $\sigma =1$). Due to this incipient crossover, 
it is difficult to identify a well defined preasymptotic regime. 
Then, in order to show its presence we show in
Fig.~\ref{lt_COP_sigma0.1_beta}, for a favorable case with small $\sigma$, i.e. $\sigma =0.1$,
how it emerges by lowering the temperature. What we see is that the curves for very small $T$
tend to become straight lines in this log-linear plot, signaling a logarithmic growth of $L(t)$. 
Regarding the intermediate regime, this is too elusive to be clearly recognized. However, as shown in the inset
of Fig.~\ref{lt_COP_sigma0.1_beta}, by plotting the quantity $\Lambda ^{-1}$
(actually we plot $(\beta \Lambda )^{-1}$ to better compare curves at different $T$) against $\tau$, as already done
in Fig.~\ref{cop_formula_Paolo}, one sees that at large times (i.e. large $\tau$) the curves tend
to have a linear behavior, which would signal the setting in of the intermediate regime, Eq.~(\ref{eq.mipoCOP}).

\bfi
 \centering
 \includegraphics[width=0.8\textwidth,clip=true]{fig_Lt_pw_COP.eps}
\caption{ {\bf (COP pow)}
$L(t)$ for a quench from $T_i=\infty$ to $T=0.4$, for $N=10^4$ and different values of
 $\sigma$, see key. The dashed green line is the $t^{1/3}$ law.}
\label{fig_Lt_pw_COP}
\efi

\bfi
 \centering
\includegraphics[width=0.8\textwidth,clip=true]{lt_COP_sigma01_beta.eps}
\caption{ {\bf (COP pow)}
$L(t)$ for a quench from $T_i=\infty$ to different values of $\beta=1/T$
(see key) for $\sigma =0.1$ and $N=10^4$. The plot is on a log-linear scale, hence
a logarithmic regime is more and more visible with increasing $\beta$.
In the inset the quantity $\Lambda ^{-1} $ is plotted against $\tau$ (see discussion around 
Fig.~\ref{cop_formula_Paolo}). Missing points in the region of relatively short times is an effect due
to the large time jumps introduced by the rejection free simulation scheme (see Appendix \ref{Appendix}).
}
\label{lt_COP_sigma0.1_beta}
\efi

\section{Deterministic continuum models} 
\label{sec_continuum}

 Although this is not the main focus of this paper, in this section we briefly discuss the 
 growth laws found in deterministic continuous models for growth kinetics. 
 This will allow us to compare the behaviour of these models
 with the ones of the Ising system analysed insofar and to discuss to which extent
 the two approaches can be considered equivalent. 
 
 For nn interactions with NCOP the kinetics can
 be described by means of the Time-Dependent Ginzburg-Landau equation,
\be
\partial_t\phi(x,t) = 2\partial_{xx}\phi -4 \phi^3 +4\phi .
\label{eq.tdgl}
\ee
This equation has time independent, single-kink solutions $\phi(x) = \pm\tanh(x)$. 
In the presence of an exponential coupling $J\ss{exp}(r)=e^{-r/R}$,
one can explicitly take into account the scale $R$ of the interactions
obtaining $\phi(x) =\pm\tanh(x/R)$. 
Combining two of such solutions
to reproduce the single domain configuration plotted in Fig.~\ref{fig.models}(a) one has
\be
\phi(x,t) = \tanh\left(\frac{x-X/2}{R}\right) - \tanh\left(\frac{x+X/2}{R}\right) + 1 .
\label{eq.kinkR}
\ee

The resulting time evolution of $X(t)$ is given by~\cite{libro}
$\dot X(t) = -\left[ V(\phi(x=0)) - V(\phi(x=\infty)) \right]$,
where $V(\phi)=(\phi^2 -1)^2$ is the standard double well potential.
Using Eq.~(\ref{eq.kinkR}) we obtain
\be
\dot X(t) =  -\left[ V(1-2\tanh(X/2R)) - V(1) \right] \equiv  -v(X/R) ,
\label{eq.6.20}
\ee
where the drift $v(X/R)$ is a positive function vanishing for small and large argument 
and with a maximum for $X/R=\tanh^{-1}(1/2)$.
In the limit $X(0)=L\gg R$, because of the exponential tails (with respect to the asymptotic values $\pm 1$) 
$v(X/R)\simeq e^{-X/R}$, which explains the logarithmic coarsening of the slow regime~\cite{NagaiKaw1986}. 
The opposite limit $X(0)=L \ll R$ (for which $v(X/R)\approx (X/R)^2$) is not physically relevant because
the two DWs model is not applicable.
However, in the intermediate regime $X(t) \approx R$ the function $v(X/R)$ 
is approximately constant because of the maximum, and
such constant drift originates a ballistic
behavior. This shows that both the slow and the ballistic regimes observed
in the Ising model have a counterpart in the continuum theory.

For the algebraic coupling $J\ss{pow}(r)$, a continuum model has been extensively studied by
Alan Bray and Andrew D. Rutenberg~\cite{BrayRut94,RutBray94},
finding $L(t) \sim t^{1/(1+\sigma)}$ for NCOP 
\footnote{In this respect it is worth stressing that such slow regime is asymptotic for $\sigma < 1$, while it is replaced by
the diffusive one ($L(t)\sim t^{1/2}$) for large $t$ if $\sigma \ge 1$. This is simply due to the relevance of temperature. In $d=1$, indeed, the model in equilibrium has a finite $T_c$ for $\sigma<1$ and a vanishing $T_c$
for $\sigma \ge 1$. Therefore, in a renormalization group language, $T$ is an irrelevant parameter in the former case and a relevant one in the latter.
If it is irrelevant, the (continuum) result found at $T=0$ is valid also switching on noise (temperature);
if it is relevant, the effect of $T$ should be visible at large enough length scales
(as it is for the short range model).}, which is the slow regime observed also
in the discrete Ising model. This regime has the same origin of the slow regime for the exponential coupling. In particular, the growth law in these regimes is a direct manifestation
of the analytic form of the interaction: it is logarithmic for an exponential $J(r)$ and power-law for
an algebraic $J(r)$.
In conclusion, for NCOP there is a rather general correspondence between the behavior of the discrete
Ising model and continuum approaches. 

For COP there are not many available results for deterministic continuum theories, 
the only available result concerning the short-range
model~\cite{Langer1971}, where a logarithmic coarsening is inferred.
Despite this, we argue on general grounds that the good correspondence between 
noiseless continuum approaches and the discrete model found for NCOP
does not apply to the conserved case. 
We say this because any regime in the Ising model with COP is
intrinsically stochastic and therefore cannot be captured by a deterministic continuum theory. 
For this reason, even if with COP a logarithmic growth is found both in the continuum 
approach~\cite{Langer1971} and in the Ising model,  these laws do not have a 
similar origin, and the corresponding regimes are physically different.
In fact, the logarithmic stage in the continuum model is due (similarly to what discussed
above for NCOP) to
an exponentially small interaction between DWs. Instead in the
Ising model it stems from the exponential vanishing of the probability that a particle detached from a 
domain reaches another cluster (because of the backward drift). 

This conclusion is corroborated by the fact that the short-time logarithmic regime in the Ising COP is 
completely general and independent of the details of $J(r)$. On the contrary, the
interaction between DWs in the continuum theory depends on the form
of the interaction, so that for an algebraically decaying coupling one does
not expect such logarithmic coarsening.

\section{Conclusions} \label{concl}

Despite that phase ordering is an old problem, a thorough
investigation of the kinetics of the Ising model with space decaying
interactions was not pursued previously.
In this paper we have considered this model in one dimension with a
coupling $J(r)$ between two spins at distance $r$ which
is a general, positive and decreasing function $J(r) > 0, \quad \forall r$.

In the asymptotic regime, $t\to\infty$, coarsening dynamics is the same as for the short range model
if $J(r)$ decays faster than $1/r^2$, the condition to have a vanishing Curie temperature
for the equilibrium model. Hence it is 
$L(t)\sim t^{1/2}$ for NCOP and $L(t)\sim t^{1/3}$ for COP.
However, our investigation of the dynamics at any time shows that
the model with space decaying interactions displays a rich and unexpected variety of different regimes,
with the asymptotic one being sometimes so delayed to be hardly observable.

Some features of coarsening dynamics are worth of note.
The first one is that extending the range of the interactions produces an
acceleration of the NCOP dynamics and a slowdown of the COP one.
Surprisingly, these opposite effects originate from the same phenomenon:
the diffusion of a DW  in the NCOP models and the diffusion of a particle in the COP models are no more symmetric,
because a drift appears.
In the nonconserved case this drift tends to move a DW towards its closest neighbour, therefore
favoring the closure of domains and speeding up the dynamics.
In the conserved case, instead, the drift is applied to a detached monomer
and reduces its possibility to attain the other clusters, therefore hampering mass exchange between
clusters and impeding the kinetics.

A second feature is related to the comparison between the dynamics of the discrete Ising model 
and the dynamics of continuum models, which has been discussed in the previous Section.
We limit here to stress that we have noticed the existence of a ballistic regime (constant drift)
in the continuum model as well. Such regime appears when we pass from the nn to the exponential model,
with a sufficiently large $R$. We have also pointed out that a deterministic
continuum approach fails in reproducing the kinetics of the Ising model
with COP in 1d, arguing that a stochastic model is needed to reproduce its
behavior in any dynamical regime, from the early stage to the asymptotic one.

In this manuscript we have provided a detailed study of the time behaviour of $L(t)$,
but this quantity does not cover all features of coarsening dynamics, 
which is also characterized by correlation functions and by the full size distribution of domains.
The general time dependent spin-correlation function $\langle s_i(t)s_j(t_w)\rangle$
is usually studied at the same site, $C(t,t_w)=\langle s_i(t)s_i(t_w)\rangle$,
or at the same time, $G(r,t)=\langle s_i(t)s_{i+r}(t)\rangle $. 
The autocorrelation function, whose scaling form is $C(t,t_w)=f(L(t)/L(t_w))$, has recently been studied for nonconserved dynamics by
the same authors~\cite{JSTAT}, discovering a new universality class appearing in the power model
when $\sigma\le 1$ (but in contrast to the dynamical exponent $z$ the function $f(x)$ and in particular the 
Fisher-Huse exponent $\lambda$, $f(x) \approx x^{-\lambda}$ for $x\gg 1$, does not depend on $\sigma$ when
$\sigma \le 1$).
The equal time correlation function is the standard spin-spin correlation function whose scaling form is 
$G(r,t)= G(r/L(t)$ and whose behavior is clearly related to the distribution of domain lengths,
$n_d(\ell,t)$. It would be of interest to study both $G$ and $n_d$ and finding out
possible connections between the two, within different models.

Beyond the case of a quench from a disordered state addressed in this paper, many other topics
remains unexplored, as for instance the kinetics following a quench from a critical 
state~\cite{Humayun_1991,PhysRevE.93.052105}, 
which is present in the one-dimensional model with algebraic interactions 
when $0<\sigma \le 1$ when $T_c$ is finite. 
In addition, besides the determination of the growth law $L(t)$, 
several other features of the Ising model with space decaying interactions are worth of further investigations. Let us mention here the aging properties, i.e. the behavior of two-time
quantities such as correlation and response functions, whose understanding
could provide useful hints for a general interpretation of aging systems~\cite{BCKM97}.
Furthermore, our studies can be extended to higher dimension $d>1$, some results for $d=2$ 
being contained in~\cite{EPL,CMJ19}, and to $\sigma < 0$,
where additivity is lost~\cite{review_long_range}.
Another interesting point to be investigated is the robustness of our results with respect to the presence of quenched disorder, which is often unavoidable in real systems. 
It is well known, in fact, that even a tiny amount of such randomness may change
radically the kinetics of coarsening systems~\cite{Corberi_crp}, both in
one dimension~\cite{Corberi_rf} and higher dimensions ~\cite{LMPZ10,CLMPZ11,CLMPZ12}.
The situation becomes even more complex if disorder introduces frustration~\cite{CMPL17,CMLP19}. In 
particular, in the case $\sigma\le 0$,   
the one-dimensional Edwards-Anderson model with algebraically decaying coupling constants, 
shows different behaviors depending on the exponent $\sigma$~\cite{KAS83,Moo86,Leu99,LS10}. 
It would be therefore interesting to explore if and to which extent the formalism developed in this paper 
may provide some hints for the understanding disordered systems with or without frustration.

A final comment concerns a recent preprint~\cite{1901.01756} whose focus is
the study of the NCOP coarsening dynamics for a lattice model with a continuum
local variable $q_i$  and falling in the equilibrium 
universality class of the Ising model with power-law interactions.
The system is a chain of oscillators with the standard single site, double-well
potential and an interaction potentials decaying as $1/r_{ij}^{1+\sigma}$,
with $0 \le \sigma \le 1$. Authors want to analyze if there is equivalence
between canonical and microcanonical ensemble or not. The answer is negative:
in the former case they obtain $z=1+\sigma$, in agreement with our results and
with those in~\cite{BrayRut94,RutBray94}. Instead, in the latter case they asymptotically find
$z=2$, showing an out-of-equilibrium ensemble inequivalence for $\sigma$-values where
additivity (and therefore equilibrium additivity) holds.

\appendix

\section{Algorithm S2 with a reduced number of interacting kinks}
\label{Appendix}

We indicate with $n$ the total number of interfaces present in the system 
at a generic time $t$  and with $x_k$ the position of the $k$-th interface (with $k=[-n/2,...,0,...,n/2]$). Under the assumption of periodic boundary conditions it is easy to show that, if $ i \in [x_k+1,x_{k+1}]$ and  $j \in [x_{m}+1,x_{m+1}]$,
then $s_i(t)s_j(t)=1$ for $k-m$ even whereas $s_i(t)s_j(t)=-1$ for $k-m$ odd. As a consequence  the Hamiltonina can be written as
 \be
H=-\sum_{k=-n/2}^{n/2}\sum_{m=-n/2}^{n/2}(-1)^{k-m} \sum_{i=x_k+1}^{x_{k+1}}
\sum_{j=x_m+1}^{x_{m+1}}J(|i-j|),
\label{eq.HDWs}
\ee
where $|i-j|$ indicate the spatial distance between the sites $(i,j)$. 
Because of periodic boundary conditions, such distance is actually given by the minimum between 
$|i-j|$ and $N-|i-j|$, but we use the expression $J(|i-j|)$ for any pair $(i,j)$ to
avoid overloading the notation. Furthermore, we assume $J(0)=0$.

We next consider the energetic cost due to the flip of the $i$-th spin 
\be
\Delta E=2 s_i(t) \sum_j J(|i-j|)s_j(t) 
\label{dde}
\ee
and separate the discussion for the NCOP and COP dynamics.

\subsection{Fast NCOP dynamics }

In our fast simulation protocol, flips of spins in the bulk are forbidden. 
Therefore only  spins at the interface ($i=x_k$ or $i=x_k+1$) can flip 
and the dynamics is mapped to the displacement of a DW 
as in Fig.\ref{fig.moves}a (upper panel) ($x_k \to x_k\pm 1$).
 If $x_k=x_{k+1}-1$, the displacement of the $k$-th interface towards the right leads to the annihilation of the two interfaces ($n \to n-2$). Annihilation also occurs for a displacement towards the left if $x_k=x_{k-1}+1$.
Without lack of generality we consider the dynamics of the defect $x_0$ and a displacement towards the right $x_0 \to x_0+1$ corresponds to the flip  of the spin $i=x_0+1$. 
According to Eqs.~(\ref{eq.HDWs},\ref{dde}) we find
\be
\Delta E  = 2 \sum_{k=-n/2}^{n/2}(-1)^k \sum_{j=x_k+1}^{x_{k+1}} J(|x_0+1-j|) .
\label{eq.HDWs1}
\ee
Similarly, a displacement towards the left, $x_0 \to x_0-1$, 
corresponds to the flip  of the spin $i=x_0$ and since, by definition $s_{x_0}=-s_{x_0+1}$ we immendiately obtain   
\be
\Delta E  = - 2 \sum_{k=-n/2}^{n/2}(-1)^k \sum_{j=x_k+1}^{x_{k+1}} J(|x_0-j|) .
\label{eq.HDWs2}
\ee
Defining 
\be
Q(|x_k-i|)=\sum_{j=x_k+1}^{x_{k+1}} J(|i-j|)
\ee
we finally obtain Eq.(\ref{deltaEii}). 

In the approximation scheme S2, we randomly choose one of the $(n+1)$ interfaces 
present at time $t$ and accept its move towards the left or the right 
with a probability given in Eq.(\ref{glauber}). 
A Monte Carlo step corresponds to $n+1$ trials. 
In particular, the approximation scheme 
with a finite number of interacting kinks
corresponds to the substitution $n\to n_K$ in Eq.~(\ref{eq.HDWs2}), where 
$n_K$ is a tunable parameter to be optimised.

\subsection{Fast COP dynamics }

We consider the Kawasaki dynamics which corresponds to the exchange of two nearest-neighbor opposite spins. With this dynamics the only possible moves  are the three listed in Fig.~\ref{fig.moves}b. 
More precisely, diffusion (upper panel) and annihilation (central panel) correspond to the simultaneous motion of two consecutive defects $x_k+1=x_{k+1}$  towards the right ($x_{k+1} \to x_{k+1}+1$, $x_k \to x_k+1$) or towards the left ($x_k \to x_k-1$, $x_{k+1} \to x_{k+1}-1$). In the annihilation process, in particular, the displacement towards the right occurs when a defect is present in the position $x_{k+2}=x_{k+1}+1$ leading to the annihilation of two defects. The same situation occurs for a displacement towards the left when $x_{k-1}=x_k-1$.
Finally the detachment process (lower panel) corresponds to the nucleation of two new DWs close to an existing one. 

The energy difference can be still obtained from Eqs.~(\ref{eq.HDWs1},\ref{eq.HDWs2}) after taking into account that  all moves involve the simultaneous flip of two consecutive spins $i=x_0$ and $i+1=x_1=x_0+1$. As a consequence  each move has an energy cost 
\be
\Delta E  = 2 \sum_{k=-n/2}^{n/2}(-1)^k \sum_{j=x_k+1}^{x_{k+1}}
\left (J(|x_1-x_k|)-J(|x_0-x_k|) \right)
\ee
which leads to 
\be
\Delta E  = 
2 J(1) -
2 \sum_{k=2}^{n/2}(-1)^k J(|x_1-x_k|)  +2 \sum_{k=1}^{n/2}(-1)^k J(|x_0-x_{-k}|)  ,
\label{eq.HDWsCOP}
\ee
corresponding  to Eq.(\ref{deltaEii}) with 
$Q\left(\vert x_j-i\vert\right)=J\left(r=\vert x_j-i \vert\right)$.

Also in the case of COP dynamics we have considered the approximation scheme with a finite number
$n_K$ of interacting kinks. Simulations have been performed according to a rejection free algorithm \cite{Bortz75}. At a given time $t$, a generic defect
$k$ can perform only one of the three moves in Fig.~\ref{fig.moves}b, with a probability $W_k=1/(1+\exp(\beta \Delta E))$ 
(see Eq.~(\ref{glauber})). At this time $t$, we select one of the defects (say the $k$-th) among the $n$ existing, with a probability $W_k/\sum_{j=-n/2+1}^{n/2} W_j$. The corresponding move is always accepted and time is incremented by $1/W_k$. Notice that, at sufficiently low temperature, after a transient the system will reach the configuration where only detachment of the kind of Fig.~\ref{fig.moves}b (lower panel) are possible. Since in this case $(\Delta E)_{nn}>0$ this moves produce huge temporal jumps ($1/W_k \gg 1$) which are, for instance, clearly visible in 
Fig.~\ref{lt_COP_sigma0.1_beta} by increasing $\beta$.    
 
\vskip 2cm\noindent
{\bf Compliance with Ethical Standards}

\noindent
Funding: F.C. acknowledges financial support by MIUR project PRIN2015K7KK8L.\newline
Conflict of Interest: The authors declare that they have no conflict of interest.\newline
Research involving Human Participants and/or Animals: It does not apply.\newline
Informed consent: It does not apply.


\begin{thebibliography}{10}
\providecommand{\url}[1]{{#1}}
\providecommand{\urlprefix}{URL }
\expandafter\ifx\csname urlstyle\endcsname\relax
  \providecommand{\doi}[1]{DOI \discretionary{}{}{}#1}\else
  \providecommand{\doi}{DOI \discretionary{}{}{}\begingroup
  \urlstyle{rm}\Url}\fi

\bibitem{Bray94}
A.~Bray, Advances in Physics \textbf{43}(3), 357 (1994).
\newblock \urlprefix\url{https://doi.org/10.1080/00018739400101505}

\bibitem{BrayRut94}
A.J. Bray, A.D. Rutenberg, Phys. Rev. E \textbf{49}, R27 (1994).
\newblock \doi{10.1103/PhysRevE.49.R27}

\bibitem{RutBray94}
A.D. Rutenberg, A.J. Bray, Phys. Rev. E \textbf{50}, 1900 (1994).
\newblock \doi{10.1103/PhysRevE.50.1900}

\bibitem{Glauber1963}
R.J. Glauber, Journal of mathematical physics \textbf{4}(2), 294 (1963)

\bibitem{Cornell1991}
S.J. Cornell, K.~Kaski, R.B. Stinchcombe, Physical Review B \textbf{44}(22),
  12263 (1991)

\bibitem{EPL}
F.~Corberi, E.~Lippiello, P.~Politi, EPL (Europhysics Letters) \textbf{119}(2),
  26005 (2017)

\bibitem{1901.01756}
F.~Staniscia, R.~Bachelard, T.~Dauxois, G.D. Ninno.
\newblock Differences in the scaling laws of canonical and microcanonical
  coarsening dynamics (arXiv:1901.01756)

\bibitem{review_long_range}
A.~Campa, T.~Dauxois, S.~Ruffo, Physics Reports \textbf{480}(3-6), 57 (2009)

\bibitem{Peierls1934}
R.~Peierls, Helv. Phys. Acta \textbf{7}(2), 81 (1934)

\bibitem{Dyson1969}
F.J. Dyson, Communications in Mathematical Physics \textbf{12}(2), 91 (1969)

\bibitem{Frohlich1982}
J.~Fr{\"o}hlich, T.~Spencer, Communications in Mathematical Physics
  \textbf{84}(1), 87 (1982)

\bibitem{Imbrie1988}
J.~Imbrie, C.~Newman, Communications in mathematical physics \textbf{118}(2),
  303 (1988)

\bibitem{Luijten2001}
E.~Luijten, H.~Messingfeld, Phys. Rev. Lett. \textbf{86}, 5305 (2001).
\newblock \doi{10.1103/PhysRevLett.86.5305}

\bibitem{Mukamel2009}
D.~Mukamel.
\newblock Notes on the statistical mechanics of systems with long-range
  interactions (arXiv:0905.1457)

\bibitem{L_asint}
G.~De~Smedt, C.~Godreche, J.~Luck, The European Physical Journal B-Condensed
  Matter and Complex Systems \textbf{32}(2), 215 (2003)

\bibitem{met_num_LR}
S.~Gupta, M.~Potters, S.~Ruffo, Phys. Rev. E \textbf{85}, 066201 (2012).
\newblock \doi{10.1103/PhysRevE.85.066201}

\bibitem{Bortz75}
A.~Bortz, J. Comput. Phys. \textbf{17}, 10 (1975)

\bibitem{Tc_sigma}
Y.~Tomita, Journal of the Physical Society of Japan \textbf{78}(1), 014002
  (2009)

\bibitem{bookRedner}
S.~Redner, \emph{A guide to first-passage processes} (Cambridge University
  Press, 2001)

\bibitem{Nato_Rodi}
P.~Politi, J.~Villain, in \emph{Surface Diffusion} (Springer, 1997), pp.
  177--189

\bibitem{libro}
R.~Livi, P.~Politi, \emph{Nonequilibrium statistical physics: a modern
  perspective} (Cambridge University Press, 2017)

\bibitem{NagaiKaw1986}
T.~Nagai, K.~Kawasaki, Physica A: Statistical Mechanics and its Applications
  \textbf{134}(3), 483  (1986).
\newblock \doi{https://doi.org/10.1016/0378-4371(86)90013-0}

\bibitem{Langer1971}
J.~Langer, Annals of Physics \textbf{65}(1), 53 (1971)

\bibitem{JSTAT}
P.P. Federico~Corberi, Eugenio~Lippiello.
\newblock Universality in the time correlations of the long-range 1d ising
  model (arXiv:1904.05595)

\bibitem{Humayun_1991}
K.~Humayun, A.J. Bray, Journal of Physics A: Mathematical and General
  \textbf{24}(8), 1915 (1991).
\newblock \doi{10.1088/0305-4470/24/8/030}

\bibitem{PhysRevE.93.052105}
F.~Corberi, R.~Villavicencio-Sanchez, Phys. Rev. E \textbf{93}, 052105 (2016).
\newblock \doi{10.1103/PhysRevE.93.052105}.
\newblock \urlprefix\url{https://link.aps.org/doi/10.1103/PhysRevE.93.052105}

\bibitem{CMJ19}
H.~Christiansen, S.~Majumder, W.~Janke, Phys. Rev. E \textbf{99}, 011301
  (2019).
\newblock \doi{10.1103/PhysRevE.99.011301}.
\newblock \urlprefix\url{https://link.aps.org/doi/10.1103/PhysRevE.99.011301}

\bibitem{Corberi_crp}
F.~Corberi, Comptes rendus Physique \textbf{16}(1), 332 (2015)

\bibitem{Corberi_rf}
F.~Corberi, A.~de~Candia, E.~Lippiello, M.~Zannetti, Physical Review E
  \textbf{65}(1), 046114 (2002)

\bibitem{LMPZ10}
E.~Lippiello, A.~Mukherjee, S.~Puri, M.~Zannetti, {EPL} (Europhysics Letters)
  \textbf{90}(4), 46006 (2010).
\newblock \doi{10.1209/0295-5075/90/46006}

\bibitem{CLMPZ11}
F.~Corberi, E.~Lippiello, A.~Mukherjee, S.~Puri, M.~Zannetti, Journal of
  Statistical Mechanics: Theory and Experiment \textbf{2011}(03), P03016
  (2011).
\newblock \doi{10.1088/1742-5468/2011/03/p03016}

\bibitem{CLMPZ12}
F.~Corberi, E.~Lippiello, A.~Mukherjee, S.~Puri, M.~Zannetti, Phys. Rev. E
  \textbf{85}, 021141 (2012).
\newblock \doi{10.1103/PhysRevE.85.021141}.
\newblock \urlprefix\url{https://link.aps.org/doi/10.1103/PhysRevE.85.021141}

\bibitem{CMPL17}
F.~Corberi, M.~Kumar, S.~Puri, E.~Lippiello, Phys. Rev. E \textbf{95}, 062136
  (2017).
\newblock \doi{10.1103/PhysRevE.95.062136}.
\newblock \urlprefix\url{https://link.aps.org/doi/10.1103/PhysRevE.95.062136}

\bibitem{CMLP19}
F.~Corberi, M.~Kumar, E.~Lippiello, S.~Puri, Phys. Rev. E \textbf{99}, 012131
  (2019).
\newblock \doi{10.1103/PhysRevE.99.012131}.
\newblock \urlprefix\url{https://link.aps.org/doi/10.1103/PhysRevE.99.012131}

\bibitem{KAS83}
G.~Kotliar, P.W. Anderson, D.L. Stein, Phys. Rev. B \textbf{27}, 602 (1983).
\newblock \doi{10.1103/PhysRevB.27.602}.
\newblock \urlprefix\url{https://link.aps.org/doi/10.1103/PhysRevB.27.602}

\bibitem{Moo86}
M.A. Moore, Journal of Physics A: Mathematical and General \textbf{19}(4), L211
  (1986).
\newblock \doi{10.1088/0305-4470/19/4/008}

\bibitem{Leu99}
L.~Leuzzi, Journal of Physics A: Mathematical and General \textbf{32}(8), 1417
  (1999).
\newblock \doi{10.1088/0305-4470/32/8/010}

\bibitem{LS10}
E.~Lippiello, A.~Sarracino, {EPL} (Europhysics Letters) \textbf{90}(6), 60001
  (2010).
\newblock \doi{10.1209/0295-5075/90/60001}

\end{thebibliography}

\end{document}